\documentclass[fleqn,usenatbib]{mnras}

\usepackage{newtxtext,newtxmath}

\usepackage[T1]{fontenc}

\DeclareRobustCommand{\VAN}[3]{#2}
\let\VANthebibliography\thebibliography
\def\thebibliography{\DeclareRobustCommand{\VAN}[3]{##3}\VANthebibliography}

\usepackage{graphicx}	
\usepackage{amsmath}	
\usepackage{amssymb}	

\usepackage{scalerel}
\usepackage{tikz}
\usetikzlibrary{svg.path}

\definecolor{orcidlogocol}{HTML}{A6CE39}
\tikzset{
  orcidlogo/.pic={
    \fill[orcidlogocol] svg{M256,128c0,70.7-57.3,128-128,128C57.3,256,0,198.7,0,128C0,57.3,57.3,0,128,0C198.7,0,256,57.3,256,128z};
    \fill[white] svg{M86.3,186.2H70.9V79.1h15.4v48.4V186.2z}
                 svg{M108.9,79.1h41.6c39.6,0,57,28.3,57,53.6c0,27.5-21.5,53.6-56.8,53.6h-41.8V79.1z M124.3,172.4h24.5c34.9,0,42.9-26.5,42.9-39.7c0-21.5-13.7-39.7-43.7-39.7h-23.7V172.4z}
                 svg{M88.7,56.8c0,5.5-4.5,10.1-10.1,10.1c-5.6,0-10.1-4.6-10.1-10.1c0-5.6,4.5-10.1,10.1-10.1C84.2,46.7,88.7,51.3,88.7,56.8z};
  }
}

\newcommand\orcidicon[1]{\href{https://orcid.org/#1}{\mbox{\scalerel*{
\begin{tikzpicture}[yscale=-1,transform shape]
\pic{orcidlogo};
\end{tikzpicture}
}{|}}}}

\usepackage{hyperref} 

\title[Solar Models: FYS]{Constraints on the Early Luminosity History of the Sun: Applications to the Faint Young Sun Problem}

\author[Basinger et al.]{
Connor Basinger,$^{1}$\thanks{E-mail: basinger.101@osu.edu}
Marc Pinsonneault \orcidicon{0000-0002-7549-7766},$^{1}$
Sandra T. Bastelberger \orcidicon{0000-0003-2052-3442},$^{2}$
B. Scott Gaudi \orcidicon{0000-0003-0395-9869}$^{1}$ 
\newauthor
and Shawn D. Domagal-Goldman \orcidicon{0000-0003-0354-9325}$^{3,4,5}$
\\
$^{1}$Department of Astronomy, The Ohio State University, 140 West 18th Avenue, Columbus, OH 43210, USA\\
$^{2}$Integrated Space Science and Technology Institute, Department of Physics, American University, Washington DC\\
$^{3}$NASA Goddard Space Flight Center
8800 Greenbelt Road
Greenbelt, MD 20771, USA\\
$^{4}$NASA GSFC Sellers Exoplanet Environments Collaboration\\
$^{5}$NASA NExSS Virtual Planetary Laboratory, Seattle, WA, 98195, USA\\
}

\date{Accepted XXX. Received YYY; in original form ZZZ}

\pubyear{2024}

\begin{document}
\label{firstpage}
\pagerange{\pageref{firstpage}--\pageref{lastpage}}
\maketitle

\begin{abstract}
Stellar evolution theory predicts that the Sun was fainter in the past, which can pose difficulties for understanding Earth’s climate history. One proposed solution to this Faint Young Sun problem is a more luminous Sun in the past.
In this paper, we address the robustness of the solar luminosity history using the YREC code to compute solar models including rotation, magnetized winds, and the associated mass loss. 
We present detailed solar models, including their evolutionary history, which are in excellent agreement with solar observables. Consistent with prior standard models, we infer a high solar metal content.
We provide predicted X-ray luminosities and rotation histories for usage in climate reconstructions and activity studies.
We find that the Sun’s luminosity deviates from the standard solar model trajectory by at most 0.5\% during the Archean (corresponding to a radiative forcing of 0.849 W m$^{-2}$). The total mass loss experienced by solar models is modest because of strong feedback between mass and angular momentum loss. We find a maximum mass loss of $1.35 \times 10^{-3} M_\odot$ since birth, at or below the level predicted by empirical estimates. The associated maximum luminosity increase falls well short of the level necessary to solve the FYS problem.
We present compilations of paleotemperature and CO$_2$ reconstructions. 1-D ``inverse'' climate models demonstrate a mismatch between the solar constant needed to reach high temperatures (e.g. 60-80 $^{\circ}$C) and the narrow range of plausible solar luminosities determined in this study. Maintaining a temperate Earth, however, is plausible given these conditions.
\end{abstract}

\begin{keywords}
Sun: activity -- Sun: evolution -- Sun: rotation -- Sun: X-rays, gamma rays
\end{keywords}



\section{Introduction}

The Faint Young Sun (FYS) problem is a longstanding problem in solar and climate modelling first introduced by \citealt{Sagan1972}. 
According to classical solar models, the Sun was 20-25\% less luminous 3.8-2.5 billion years ago (Ga) during the Archean Eon (\citealt{Bahcall2001}). Assuming otherwise present-day conditions on Earth, this would result in a surface temperature for the Earth that was too low for liquid oceans in the distant past. 
This assumption is in disagreement with geological evidence both for liquid water on the Earth's surface and for warm surface temperatures soon after Earth's formation 4.568 Ga. 
The solution to the FYS problem is of interest for understanding the early habitability of the planets in our own solar system and has implications for that of exoplanets.

The earliest indirect evidence for liquid water on Earth's surface comes through Hadean zircons (4.4-4.2 Ga, \citealt{Wilde2001}, \citealt{Mojzsis2001})
and the earliest direct evidence in pillow basalts and layered banded iron formations (3.8 Ga; \citealt{Holland1984}; \citealt{Appel2001}) 
as well as fluid inclusions in hydrothermal quartz (3.5 Ga, \citealt{Foriel2004}). 
Regarding temperature, isotopic ratios of $^{18}$O/$^{16}$O imply global mean surface temperatures of 60-80 $^{\circ}$C (\citealt{KNAUTH1976}; \citealt{Knauth2003}; \citealt{Tartese2017}). These studies and their methods remain controversial, relying on assumptions such as the Archean oceans having a similar $\delta^{18}$O to a modern, ice-free ocean (\citealt{Hren2009}; \citealt{Blake2010}; \citealt{Sengupta2020}). 
In contrast to the high temperatures implied by isotopic ratios, various glacial rocks imply global mean temperatures that fell below 20 $^{\circ}$C at times during the Archean,
with the Huronian glaciations marking the end of the Archean at 2.5 Ga (\citealt{Kasting2003}).

Several solutions to the FYS problem have been proposed (see, e.g., \citealt{Sagan1972}, \citealt{Kasting2003}, \citealt{Feulner2012}, \citealt{Charnay2020}). Increasing the atmospheric abundances of greenhouse gases such as CO$_2$ or CH$_4$ is most often invoked as a solution (\citealt{Feulner2012}). Other proposed ideas include: 
increasing atmospheric pressure to broaden absorption lines (\citealt{Goldblatt2009}), though there are claims that air pressure was lower during this time (e.g. \citealt{Som2016}); 
lowering the Earth's surface albedo, such as by constraining the emerged land fraction (\citealt{Jenkins1993}; \citealt{Rosing2010}; \citealt{Goldblatt2011b}); 
investigating changes in Archean cloud properties and distribution which could result in stronger warming in climate models (\citealt{Goldblatt2011a}; \citealt{Goldblatt2021}; \citealt{Yan2022});
and making the early Sun more luminous (\citealt{Graedel1991}; \citealt{Whitmire1995}). 
In this paper, we focus on the idea that the early Sun could have been more luminous than expected based on standard solar models. We aim to quantify the uncertainties in the Sun's history due to the input physics in the models, the adopted solar abundances, and the solar mass loss history.
This should be considered more broadly in the context of geological records that constrain atmospheric composition, surface albedo, and surface temperature.

There are sound theoretical reasons to expect the Sun to have been fainter in the past. As the Sun fuses hydrogen into helium in its core, the number of particles decreases and the mean molecular weight increases. In response, the core contracts and reaches higher temperatures in order for the pressure to maintain its balance against gravity. Energy generation is a strong function of temperature, so the Sun's luminosity increases. For the same reason, more massive stars, which have higher core temperatures, are more luminous. Some authors have argued that the Sun could have been more luminous in the past than expected based on standard solar models if it were more massive and subsequently lost that mass in a magnetized wind. Even in the absence of the solar wind, the mass of the sun will still decrease because the product of hydrogen fusion is less massive than the reactants. The difference in masses is converted to energy via $E=mc^2$, and the Sun will eventually radiate this energy via its luminosity.

Mass loss affects the flux received by the Earth from the Sun in two ways. 
First, the luminosity of a star is a strong function of the mass (\citealt{Parker1965}). These are classically related on the main sequence via a power law $L \propto M^{\alpha}$, where a typical value $\alpha \sim 3.5$ is determined through an empirical fit to observed data. The exact relationship is determined via a combination of the energy generation, which increases with mass and temperature, and the opacity, which decreases with temperature allowing for more efficient energy transport. Values of $\alpha$ from 3 to 5 are often cited in FYS papers. Near the mass of the Sun ($1.0 < M/M_\odot \le 1.1$) the power is locally higher than across the entire main sequence, with $\alpha \sim 4.5$ from empirically fitting to MIST isochrones \citep{Choi2016}. 
Second, mass loss can alter the Earth-Sun distance such that a more massive Sun was closer to Earth. Assuming slow (adiabatic) mass loss, the orbit of the Earth will increase over time such that $a_\oplus \propto M^{-1}$, and thus the flux received at the Earth will increase as $M^2$. 
The extra power of 2 for the flux results in $F \propto M^{\alpha+2}$.
Even a slightly more massive Sun could therefore dramatically increase luminosity of the Sun and the flux at Earth. Only a $\sim$6\% increase in mass would be enough to give the Sun the same luminosity 4.568 Ga as it has today \citep{Charnay2020}.
However, this would require an average mass loss rate over the main sequence lifetime of the Sun more than 600 times higher than the current rate.

There are also sound reasons to expect that the Sun experienced a higher mass loss rate in the past. The activity, mass loss, and rotation of a star are coupled due to the magnetized solar wind (e.g. \citealt{Hartmann1987}). The stellar dynamo generates the Sun's magnetic field and hot stellar coronae, which are linked to X-ray emission and mass loss. Charged particles are carried away from the Sun along magnetic field lines, conserving angular momentum and applying a torque that slows the Sun's rotation.
\cite{Skumanich1972} found that there was a relationship between stars' rotational velocities $v$ and their ages $t$ such that the velocity decreases with age as $v \propto t^{-1/2}$.
The Sun's faster rotation when it was younger corresponds to higher levels of activity and higher X-ray fluxes. Higher X-ray fluxes are correlated to higher mass loss rates in solar analogs \citep{Wood2002}. 
The majority of the Sun's mass loss occurs early in its lifetime due to its more rapid rotation during this time.

The present-day solar mass loss rate is small, $\dot{M}_\odot = 2 \times 10^{-14} M_\odot yr^{-1}$, so mass loss has traditionally been neglected in solar models. However, both observations and models predict higher mass loss rates for young stars.
Most papers in the literature find that these mass loss rates are nonetheless insufficient to explain the FYS problem. \cite{Wood2002} empirically derived mass loss rates from solar analogs and found that the Sun's total mass loss was less than 0.03 M$_\odot$ (with a best estimate of 6-7 $\times 10^{-3}$ M$_\odot$) with the majority of the mass loss occurring within the first few 100 Myr.
\cite{Johnstone2015} calculated an expected mass loss < $3 \times 10^{-4}M_\odot$ in a magnetized wind.

Others have argued in favor of greater mass loss.
For example, \cite{Martens2017} defended the plausibility of greater mass loss for a more luminous Sun. They argued for sustained mass loss in a stronger wind based on an order of magnitude estimate relating the torque due to mass loss to the angular momentum of the Sun. However, they did not perform a fully self consistent calculation including the rotation history of the Sun. In their model, the current solar mass loss rate also would induce a torque much smaller than that which is observed.
Papers which advocate for arbitrary mass loss to satisfy the FYS problem (e.g. by saying that a few percent more massive Sun could compensate for the lower luminosity of the young Sun) fail to account for the fact that the Sun's rotation history will set bounds on the allowed amount of mass loss with feedback based on the Sun's angular momentum.
This feedback - that high mass loss rates imply high angular momentum loss rates, which in turn reduce mass loss - has not had a full, self-consistent treatment in models previously used to address the FYS problem.

There is also uncertainty in the Sun's mass loss history because the birth conditions of the Sun - whether it was born as a rapid, moderate, or slow rotator - are still under debate. 
Solar mass stars arrive on the main sequence with a range of rotation rates, and the majority of them are already slow rotators, with correspondingly modest winds, by the age of the Pleiades cluster (125 Myr; \citealt{Rebull2016}; \citealt{Stauffer1998}). The origin of the range of rotation rates is a matter of active research, but the standard assumption adopted in angular momentum evolution modeling involves magnetic coupling between protostars and their accretion disks (\citealt{Koenigl1991}; \citealt{Shu1994}), enforcing a nearly constant rotation rate during the massive gaseous disk phase. A star with a short disk lifetime will be released from its disk when it is large, and it will become a rapid rotator; a star wtih a long-lived or massive disk will be released when it is smaller, producing a slow rotator. Although the detailed mechanism is controversial (\citealt{Matt2005}), something like this is needed to understand the existence of young slow rotators in star clusters.
While enhanced lithium depletion in the Sun suggests that it was born as a rapid rotator \citep{Pinsonneault2001},
another recent paper argued that a rapid rotator was disfavored to avoid a more active Sun leading to rapid escape of the atmosphere on Earth \citep{Johnstone2021}. 
A long Jupiter formation timescale \citep{Kruijer2017} may suggest a long disk locking time and thus slow or moderate early rotation.
Because a rapid rotating Sun has more activity and higher mass loss than a slow rotating Sun, the birth rotation period of the Sun is one of the largest uncertainties in the Sun's early luminosity history.

The luminosity history of the Sun could also be altered by incorporating other new physics such as including star spots, changing nuclear cross sections, or changing the composition or opacities used in the models.
Their exact impacts may be quantified through perturbing standard solar models. The accuracy with which the luminosity evolution is known changes based on the uncertainties of the input parameters. 
Past work in the literature has shown that deviant models (e.g. no diffusion) only change the Sun's luminosity by up to $\sim$1\% \citep{Bahcall2001}.
\cite{Bahcall2005} demonstrated that there were uncertainties in the calculated neutrino fluxes of the Sun due to changing the abundances of certain heavy elements. In addition, other papers have noted that changing the chemical composition of the Sun can lead to disagreements with helioseismic constraints on the Sun's structure such as the depth of the convective zone or the sound speed (e.g. \citealt{Asplund2021}).


\begin{figure}
  \centering
	\includegraphics[width=\columnwidth]{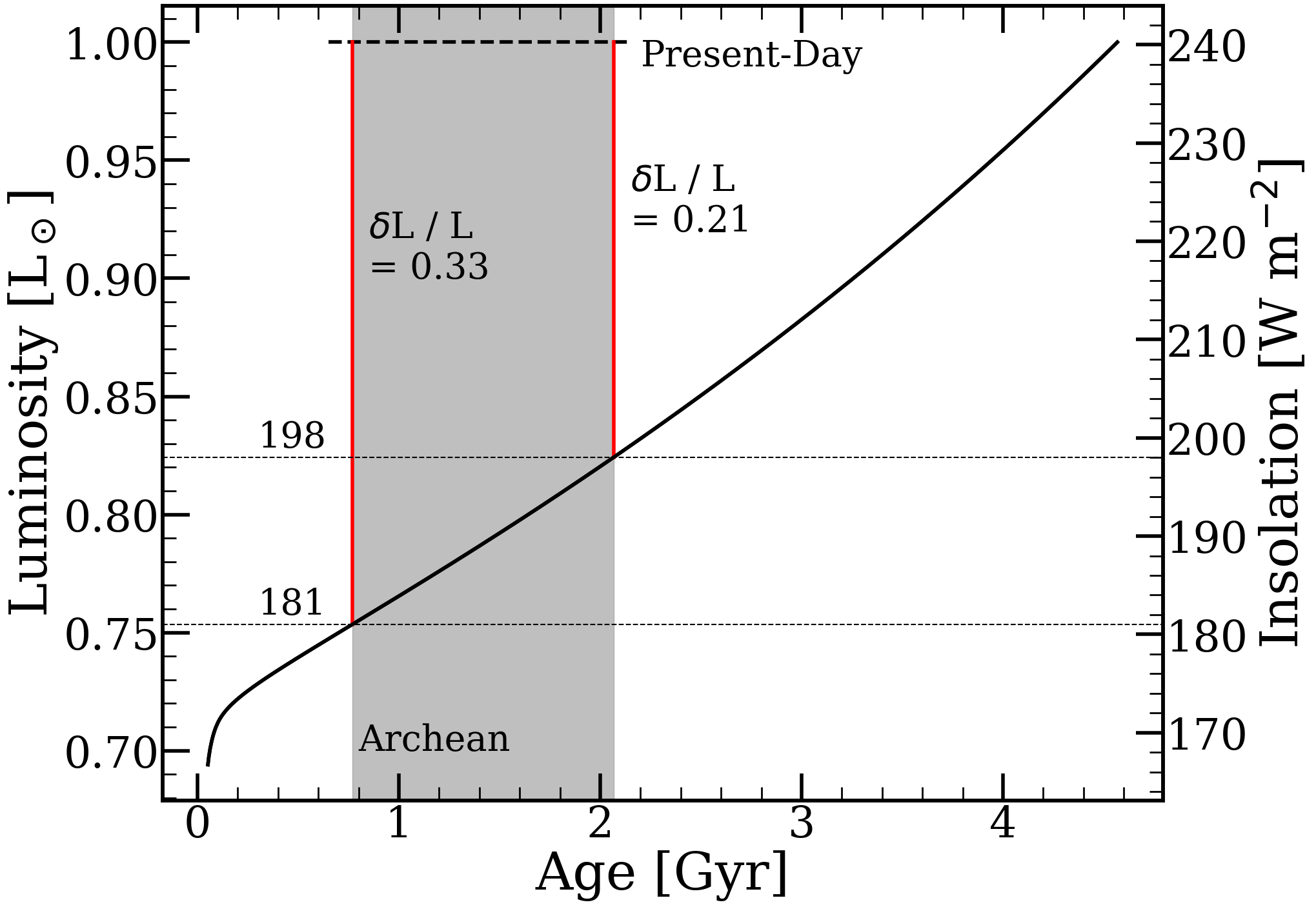}
    \caption{Time evolution of the solar luminosity and mean annual global solar insolation for our reference solar model. The horizontal dashed line shows a value of 1 L$_\odot$ (240 W m$^{-2}$) during the Archean Eon (3.8-2.5 Ga). Assuming the present-day surface temperature, albedo, and atmospheric conditions for Earth, this requires fractional increases in luminosity of $\delta$L/L = 0.21 and 0.33 at the ages identified in red. The dotted lines show the insolation at the bounds of the Archean Eon. The deficit that must be accounted for is 42-59 W m$^{-2}$ also assuming a constant semi-major axis for Earth's orbit (a scenario without mass loss).}
    \label{fig:sol_irr}
\end{figure}

Conversely, if the solar history is well constrained, we can explore the consequences for models of the ancient Earth. Such reconstructions will depend on both the time history of the Sun and that of the climate on Earth, but the constraints that emerge are nonetheless powerful. 

As a byproduct of our efforts to update classical solar models to include effects not traditionally included in solar models, we have developed an improved standard model of the evolution and interior of the sun. Improved standard solar models are of general interest to investigate the effects of refining parameter estimates or expanding the scope of the models to include additional physical effects. For example, models used for neutrino and helioseismology studies have not typically included the effects of rotation, mass loss, or magnetic fields.

In this work, we first present a non-rotating, classical solar model as a reference model in \S2. Any subsequent changes to the luminosity are measured relative to this model. In \S3, we then investigate the effects of changing the solar composition and perturbing the input physics on the solar luminosity history and to what extent this can contribute to solving the FYS problem. We construct new standard solar models including updated compositions, rotation, and for the first time, a fully self-consistent treatment of mass loss and the angular momentum evolution of the Sun. In \S4, we use reconstructions of the temperature and atmospheric composition of the Earth to investigate the surface temperatures that can be maintained based on our luminosities and the corresponding solar constant. Finally, we present the characteristics of our new standard solar model including rotation and mass loss in \S5, which we believe is the best model of the Sun to date.

\section{The Classical Solar Model}

In this section we present our classical standard solar model (e.g. \citealt{Bahcall1992}, etc.), which we adopt and henceforth refer to as our reference model. Our reference model follows the traditional structure of solar models computed for studies of helioseismology and neutrinos in order to facilitate comparison with the literature. We begin in the pre-main sequence, at the end of the deuterium burning birthline. This anchors the start of our calculations at the formation epoch of the oldest meteorites. We then evolve models to the age, luminosity and radius of the present-day Sun. We focus on the time evolution of the bolometric luminosity, the most important component for the FYS problem. 

Our reference model and all subsequent variations on this model are calibrated to the current solar age, luminosity, and radius (see Table~\ref{tbl:refer}; age of 4.568 Gyr, L$_\odot = 3.828 \times 10^{33}$ ergs/s, R$_\odot = 6.957 \times 10^{10}$ cm; \citealt{IAUB3}). We adopt a solar mass of $1.988 \times 10^{33}$g (\citealt{IAUB3}). 
We explore the uncertainties in our reference model due to uncertainties in the prior assumptions. In \S2.2 we will perturb the input physics of the models for comparison with our reference model. 
In \S3 we will add physical effects that may affect the time evolution of the Sun, but are not traditionally included in standard solar models. Examples of such effects include rotation, magnetic fields, and mass loss, and their inclusion allows us to compute our best case standard solar model. 

\subsection{Reference Model} 

We construct our reference model using the Yale Rotating Evolution Code (YREC; \citealt{vansaders2012}). Our overall approach is similar to that employed by \cite{Bahcall1992} and subsequent papers. We adopt our reference case with state of the art physics. We adopt opacities from the Opacity Project (\citealt{Badnell2005}; \citealt{Delahaye2016}) with meteoritic metal abundances from \cite{Magg2022}. We include gravitational settling and thermal diffusion of helium and heavy elements using the \cite{Thoul1994} formalism, Kurucz atmosphere and boundary conditions (\citealt{Castelli2003}), the OPAL equation of state (\citealt{Rogers1996}; \citealt{Rogers2002}), and nuclear reaction cross sections from \cite{Adelberger2011}. 
We calibrate our models by adjusting the mixing length and surface helium abundance in order to match the present-day luminosity and radius of the Sun (see Tables~\ref{tbl:conv_prop} for adopted values). Changing the mixing length primarily calibrates the radius whereas the surface helium abundance primarily calibrates the luminosity. We also require our models to match the observed surface metal to hydrogen ratio Z/X, which differs from the birth ratio because of gravitational settling and thermal diffusion (\citealt{Bahcall1992}; \citealt{Bahcall1995}).

\begin{table}
\centering
\caption{Reference Model Parameters}
\begin{tabular}{l l l}
\multicolumn{1}{c}{Parameter}
&\multicolumn{1}{c}{Value}
&\multicolumn{1}{c}{Reference}\\
\hline
Age & 4.568 Gyr & \cite{Moynier2007} \\
Luminosity & $3.8275 \times 10^{26}$ W & \cite{IAUB3}\\
Radius & $6.957 \times 10^{8}$ m & \cite{IAUB3}\\
Mass & $1.988 \times 10^{30}$ kg & \cite{IAUB3}\\
Z/X & 0.0226 & \cite{Magg2022} \\
\end{tabular}
\label{tbl:refer}
\end{table}

Fig.~\ref{fig:sol_irr} shows the evolution of the solar luminosity scaled to the current solar luminosity as a function of time predicted by our reference models and thus quantifies the magnitude of the FYS problem. 
The increase in luminosity during the main sequence is the origin of the Faint Young Sun problem, with the predicted luminosity being 18-25\% fainter at the bounds of the Archean (3.8-2.5 Ga). 
The horizontal dashed line shows a luminosity of 1 L$_\odot$, corresponding to the luminosity required for the present-day surface temperature assuming a constant albedo and atmosphere for Earth. 
Fractional changes in luminosity of $\delta$L/L = 0.33 and 0.21 at the beginning and end of the Archean are required to reach this value. In order to be sufficient to solely explain the FYS problem, the luminosity of the Sun would have to be underestimated by as much as 21-33\% in our reference model, while still reaching the same present-day value. 



Another common set of units used to describe the FYS problem instead uses the mean annual global solar insolation as shown in Fig.~\ref{fig:sol_irr}. This is distinct from the solar constant $S$, which measures the flux of the Sun at the distance of Earth. The mean annual global solar insolation averages the flux the Earth receives from the Sun over the Earth's surface area, accounting for the albedo of the Earth. It is expressed as
\begin{equation}
F = \frac{L}{4 \pi a^2} \left(\frac{\pi R_{e}^2}{4 \pi R_{e}^2}\right)(1-A_B) =
\frac{L}{16 \pi a^2} (1-A_B)
\end{equation}
where $L$ is the luminosity of the Sun, $a=1$AU is the semi-major axis for the Earth orbiting the Sun, $R_{e}$ is the radius of the Earth, and $A_B\sim0.294$ is the Bond albedo for Earth (\citealt{Wild2012}). We assume $a$ and $A_B$ to be constant for Fig.~\ref{fig:sol_irr}.
The radiative forcing is the difference in mean annual global insolation for a perturbed model and our reference model.
The present-day value of the mean annual global solar insolation is 240 W m$^{-2}$. Values of 181 W m$^{-2}$ (3.8 Ga) and 198 W m$^{-2}$ (2.5 Ga) corresponding to the bounds of the Archean show that there is a deficit of 42-59 W m$^{-2}$ for our reference model. Cumulatively, solutions to the FYS problem must contribute at this level to have maintained present-day conditions on Earth.
Moving forward, we will report the radiative forcing alongside the fractional change in luminosity for the reader's convenience.

The evolution of the solar radius and $T_{\rm eff}$ as a function of time follow a similar trend as the luminosity with a monotonic increase to the present age of the Sun. 
The present-day effective temperature is $T_{\rm eff}$ = 5772 K \citep{IAUB3}. $T_{\rm eff}$ is defined in terms of the radius and luminosity of the Sun assuming it is a blackbody, ensuring consistency between these values.

\subsection{Variations on Reference Model Inputs}

Intrinsic sources of uncertainty in the predicted luminosity evolution of the sun arise from the uncertainties in the input physics, including the equation of state, opacities of matter to light, energy generation, and energy transport. There are also measured uncertainties in values such as the nuclear reaction cross sections or the surface metal to hydrogen ratio (often referred to as Z/X, where Z is the mass fraction of all elements heavier than helium and X is the mass fraction of hydrogen). The work in this section investigates the influence that perturbing different inputs to our reference model in \S2 has on the output luminosity. Each set of models varies an individual input parameter. These models are then compared to our reference model to investigate potential uncertainties in the luminosity of the reference solar model due uncertainties in the input physics. Based on extensive solar model studies, nonlinear interactions induced by simultaneous modest changes in input physics are small, justifying our perturbation approach. We assume that changes are uncorrelated, add them in quadrature, and treat systematic errors in input physics as effective 2$\sigma$ or 3$\sigma$ uncertainties. See \cite{Bahcall1992} for a detailed justification of these assumptions. Impacts on the other aspects of the models (e.g. sound speed) are described in \S5.

\subsubsection{Changing Solar Compositions}
Our reference model in \S2.1 uses meteoritic metal abundances from \cite{Magg2022}, hereafter M22M. In this subsection, we investigate models using the classic metal abundances presented in \cite{Grevesse1998} and other recent abundances from \cite{Asplund2021} and photospheric abundances from \cite{Magg2022}, hereafter GS98, AAG21, and M22P, respectively (see Table \ref{tbl:param}).
Each set of metal abundances uses its corresponding opacity tables in our models with surface Z/X identified in Table~\ref{tbl:param}.
GS98 and AAG21 use a mix of meteoritic and photospheric abundances, while M22M adopts meteoritic abundances where available and M22P adopts photospheric abundances where available. Even for M22M, e.g., photospheric abundances must be adopted for volatile elements like C, N, and O, which are depleted in meteorites. 

For almost 20 years, there has been a persistent disagreement in the literature on the heavy element content of the Sun.
GS98 summarized the standard solar composition based on (at the time) updated modeling of new solar spectra. The authors used the empirical \cite{Holweger1974} atmospheric structure, and assumed LTE for important species. A series of changes to model inputs led to a different family of solutions, first popularized in \cite{Asplund2005}. Hydrodynamic simulations, rather than a fixed microturbulence parameter and a 1D empirical temperature structure, were used; in addition, departures from LTE and improvements in the atomic physics were included. The net result was a substantial reduction in the overall predicted heavy element content of the Sun, as seen in the highly influential \cite{Asplund2009} summary.

AAG21 revised the composition of the solar photosphere using 3D non-LTE calculations for additional elements compared to their previous work \citep{Asplund2009}. See Table 1 of AAG21. AAG21 adopted meteoritic abundances for 15 elements. 
They argued that the older stellar abundances using 1D LTE calculations like GS98 should be discarded, however, a new generation of independent model atmosphere calculations (\citealt{Caffau2008}; \citealt{Caffau2011}) including non-LTE effects and 3D model atmospheres has emerged, and the solar abundances derived from these calculations are systematically higher than to the \cite{Asplund2009} scale. This is in part tied to judgement calls concerning which features to include (see \citealt{Pinsonneault2009}). 
M22 used up-to-date 3D non-LTE calculations and simulations, derived from an independent code. M22 argued that many of the differences in elemental abundances between their method and that of AAG21 can be attributed to artefacts or data reduction, particularly for weak features. 

Modern stellar atmosphere models have therefore been used to advocate for both a higher (M22M) and lower (AAG21) abundance scale. We thus treat the solar heavy element mixture as a source of systematic error. We adopt M22M as our base case because solar models computed with this heavy element mixture are in excellent agreement with helioseismic data (see our discussion of the solar sound speed profile in \S5; Fig.~\ref{fig:csound_all}). By contrast, the lower metallicities in AAG21 lead to a much worse agreement with these data.

\begin{figure}
  \centering
	\includegraphics[width=\columnwidth]{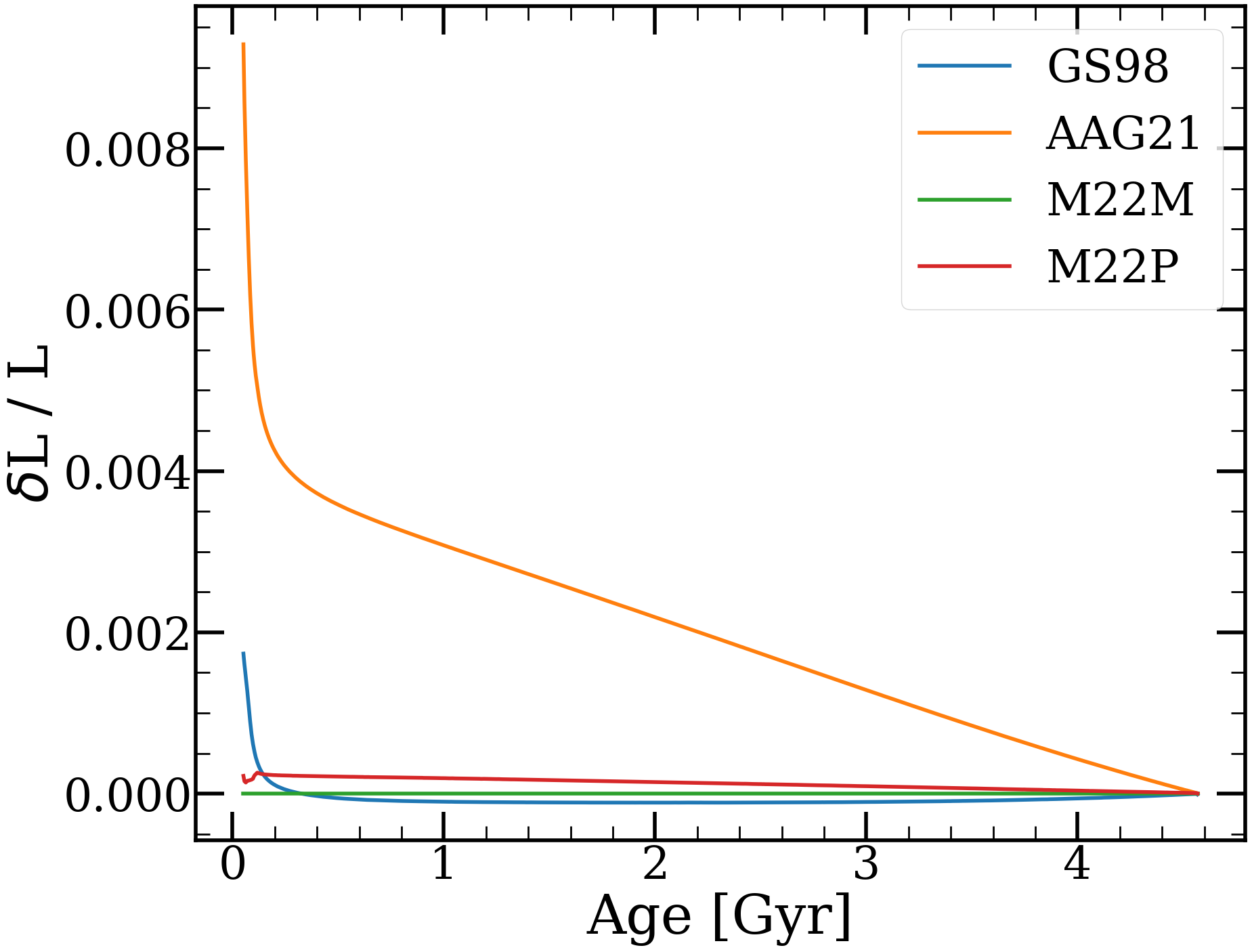}
    \caption{Fractional difference in luminosity for the GS98 (blue), AAG21 (orange), and M22P (red) abundance mixtures measured relative to M22M (green). There is generally good agreement with even AAG21, the most discrepant model, showing an increase of only a few tenths of a percent at early times.}
    \label{fig:l_diff_all}
\end{figure}

Fig.~\ref{fig:l_diff_all} shows the fractional difference in the luminosities as a function of age for the solar models with the adopted metal abundances labeled in the legend. Each luminosity is measured relative to our reference model. The fractional difference in the luminosity of the model adopting the AAG21 abundances increases approximately linearly with time before the present up to $\sim$0.4\% around $\sim$100 Myr. This fractional difference increases rapidly during the pre-main sequence because the luminosity is changing rapidly with time, and thus even small changes in the composition can lead to large fractional differences in the luminosity.  Nevertheless, we find that the fractional difference in the luminosity is still always below $\sim$1\%.
In contrast, the fractional changes for the models adopting the GS98 and M22P abundances are $<0.05$\% for essentially all of the Sun's history, with the GS98 model exhibiting the similar rapid increase prior to $\sim$100 Myr.
The choice of chemical composition can slightly alter the Sun's luminosity history. In particular, adopting lower metal abundances can lead to a higher luminosity on the order of a few tenths of a percent.
Nonetheless, with such a small luminosity change compared to that which is required, we conclude that the solar heavy element mixture which we choose to adopt in our models does not effect the Sun's luminosity at a level necessary to solve the FYS problem. 

\subsubsection{Perturbing Input Physics}

The work that follows builds upon our reference model, aiming to investigate deviations caused by the uncertainties in the input physics. We vary the individual parameters of our reference model to determine the influence of each parameter on the luminosity history.

Table \ref{tbl:param} summarizes the parameter variations and includes references for their values and uncertainties.
Historically, there were modest zero point uncertainties in the solar luminosity (0.4\% in \citealt{Bahcall1992}) and age (30 Myr). Over the past decade, the uncertainties in both have declined dramatically. 
With current data, neither is a major contributor to the error budget. 

The effects of diffusion and gravitational settling could potentially have a greater influence on the solar luminosity by, like the metal abundances, changing the interior densities, temperatures, and opacities.
The p-p and $^{14}$N-p nuclear reaction cross sections directly influence the luminosity by changing the rate of energy generation. Feedback can cause certain parameters to counter-intuitively impact the luminosity. 
At fixed temperature, a higher p-p cross section increases the luminosity of a solar model. However, a calibrated solar model has the solar luminosity fixed, so increasing the p-p cross section has the effect of lowering the model central temperature.
On the other hand, the $^{14}$N-p cross section doesn't have the same counter-intuitive impact because the CNO cycle is a small part of the Sun's energy generation. 
A higher diffusion rate increases the central metal content relative to the surface value, which is fixed in the calibration process. This leads to a higher core temperature and a greater energy generation rate and luminosity. Helium diffusion has a similar effect.

\begin{table}
\centering
\caption{Input parameters with uncertainties.}
\begin{tabular}{l l l}
\multicolumn{1}{c}{Parameter}
&\multicolumn{1}{c}{Value}
&\multicolumn{1}{c}{Reference}\\
\hline
Age & $4.568\pm0.001$ Gyr & \cite{Moynier2007} \\
Luminosity & $3.8275 \pm 0.0025 \times 10^{26}$ W & \cite{Frohlich2012} \\
Abundances & X = 0.7392 & M22M \\
& Z/X = 0.0225 & \\
& X = 0.7394 & M22P \\
& Z/X = 0.0226 & \\
& X = 0.7438 $\pm$ 0.0054 & AAG21 \\
& Z/X = 0.0187 & \\
& X = 0.735 & GS98\\
& Z/X = 0.0231 & \\
$\sigma_{pp}$ & $4.01\pm0.04\times 10^{-22}$ keV b & \cite{Adelberger2011}\\
$\sigma_{N_{14}p}$ & $1.66\pm0.12$ keV b & \cite{Adelberger2011}\\
Diffusion & $1.0\pm15$\% & \cite{Thoul1994} \\
\end{tabular}
\label{tbl:param}
\end{table}

\begin{figure}
  \centering
	\includegraphics[width=\columnwidth]{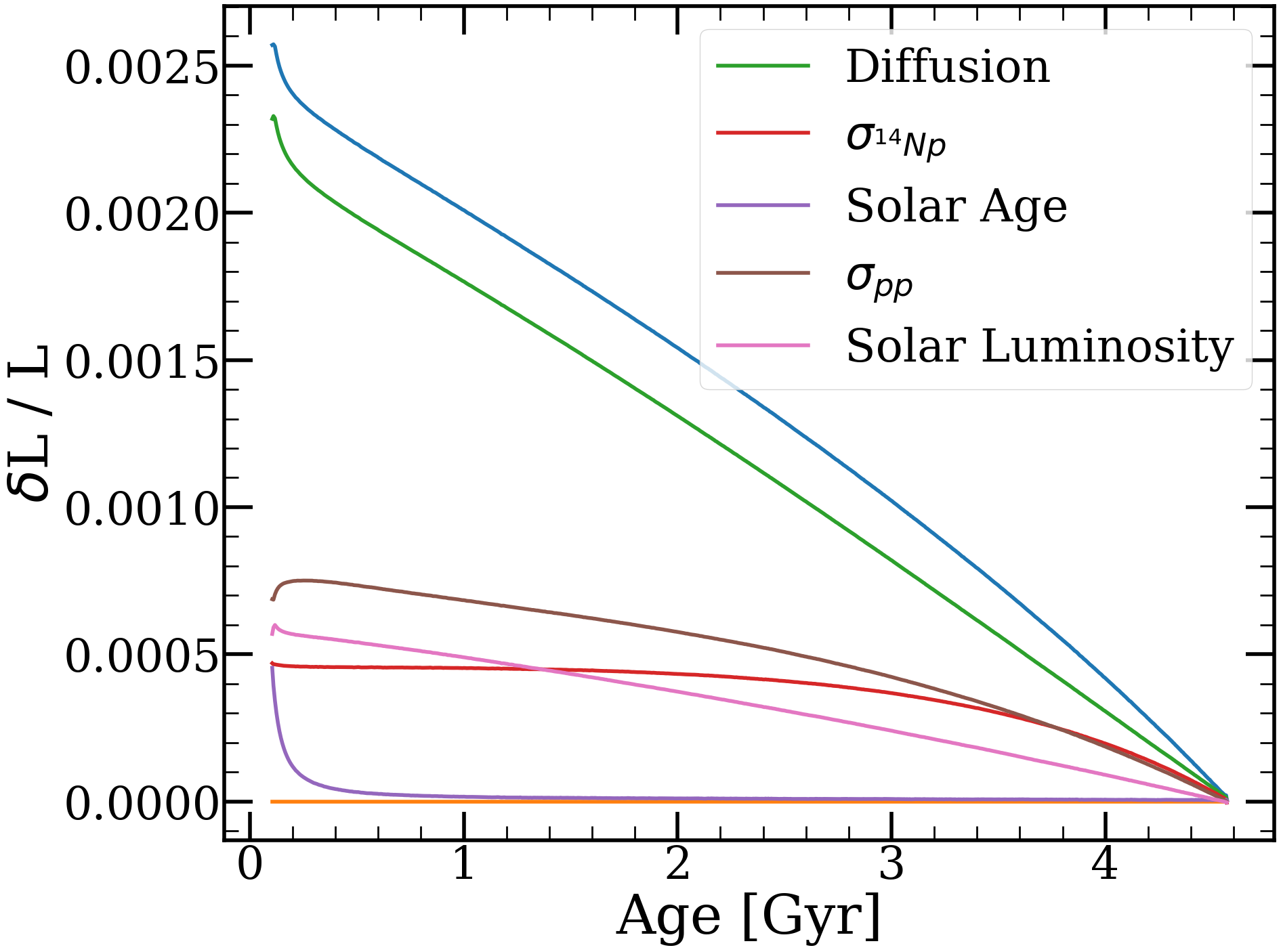}
    \caption{The fractional change in the luminosity computed from varying the input physics in Table \ref{tbl:param} by $\pm 3\sigma$ compared to our reference model. The luminosity histories that correspond to each individual parameter are identified in the legend. The individual luminosity histories are added in quadrature to compute the total positive luminosity error (blue line). The errors about the luminosity of the reference model are nearly symmetric, so it is also plausible that the luminosity could instead decrease by a comparable amount.}
    \label{fig:lum_errors}
\end{figure}

We compute the partial derivatives of the luminosity with respect to the various input parameters. These are calculated through comparing the result of a variant model with an individual parameter varied by 3$\sigma$ to our reference model. Note that diffusion is an exception which we do not vary by 3$\sigma$, but instead by the 15\% 1$\sigma$ uncertainty. These partial derivatives are a function of time with the partial derivatives equal to 0 at the present day due to the models' construction with a solar calibration. 
Table \ref{tbl:rad_forcing} presents these partial derivatives multiplied by the uncertainties, then added in quadrature and converted to the appropriate radiative forcings at 3.8 and 2.5 Ga,
the bounds of the Archean Eon when liquid water existed on Earth's surface.

Fig.~\ref{fig:lum_errors} illustrates the difference in solar luminosity compared to the standard solar model as a function of age considering variations in the input physics. The top line (blue) is the upper limit on solar luminosity. 
To obtain these uncertainties, three times the partial derivatives times the uncertainties in the parameters (illustrated by the other lines) are added in quadrature. The individual parameters are identified in the legend.
Adding the luminosities of the different models in quadrature assumes that all of these uncertainties are independent and Gaussian, which is undoubtedly an incorrect assumption. We do not know how these uncertainties are distributed. One could make an argument that adding the absolute value of each of the partial derivatives would provide an estimate of the maximum change in luminosity that is possible, but the central limit theorem suggests that our estimate should be approximately correct. 

\begin{table}
\centering
\caption{Uncertainties in radiative forcings at the bounds of the Archean.}
\begin{tabular}{c c c}
\multicolumn{1}{c}{Parameter}
&\multicolumn{1}{c}{3.8 Ga}
&\multicolumn{1}{c}{2.5 Ga}\\
\multicolumn{1}{c}{}
&\multicolumn{1}{c}{[W m$^{-2}$]}
&\multicolumn{1}{c}{[W m$^{-2}$]}\\
\hline
Total Positive & 0.382 & 0.299 \\
Total Negative & -0.376 & -0.292 \\

\end{tabular}
\label{tbl:rad_forcing}
\end{table}

The full distribution of solar luminosities is roughly symmetric about our reference model, so it is possible for uncertainties in the input parameters to lead to a decrease in the inferred luminosity instead.
This figure shows a maximum increase (or decrease) in solar luminosity of about 0.25\% at early ages, which lies between roughly the same order of magnitude to several times less than the potential change induced by the solar composition. 
As mentioned above, the central limit theorem suggests that the parameter errors may combine in such a way that the luminosity history of the Sun is not substantially different than our reference model.
Nonetheless, the case that the luminosity corresponding to each of these parameters is at a maximum is a useful upper limit on the Sun's plausible luminosity history, and thus an upper limit on the contribution of solar model inputs to the FYS problem. With a radiative forcing of 0.382-0.299 W m$^{-2}$ as an upper limit, this is insufficient to solely explain the FYS problem.

\section{Angular Momentum and Mass Loss via Magnetized Wind}

Stars like the sun with thick outer convective zones are thought to generate strong ordered magnetic fields via the dynamo mechanism (e.g. \citealt{Parker1955}; \citealt{Kraft1967}). These magnetic fields, coupled with mass loss due to an ionized stellar wind, cause these stars to spin down over time (e.g. \citealt{Weber1967}). Thus, the mass loss rates of such stars are coupled to their spin down rates.  Since a change in the mass of star leads to a change in its luminosity, the luminosity evolution can be tied to the spin down rate.  Furthermore, the magnetic fields in sunlike stars leads to X-ray emission, with the strength of this emission coupled to the strength of the magnetic field, which itself depends on the rotation rate of the star. As we have measurements of the rotation rates and X-ray emission of sunlike stars at a wide variety of ages, we can calibrate physical models of mass loss due to a magnetized wind to emprical data. 

Here we model angular momentum loss from birth to the present-day age of the Sun using the \cite{Matt2012} wind model, using the \cite{vansaders2013} model (hereafter VP13) to relate the mass loss rate and the magnetic field strength to stellar observables. We calibrate this model using observations of the rotation rate of sunlike stars in clusters of a variety of ages. 

Following VP13, the magnetic field strength scales as 
\begin{equation}
\frac{B}{B_{\odot}} = Ro^{-1} \left(\frac{P_{phot}}{P_{phot,\odot}} \right)^{1/2},
\end{equation}
where $P_{phot}$ is the photospheric pressure, and the Rossby number $Ro$ is defined as
\begin{equation}
    Ro \equiv \left( \frac{\omega}{\omega_{\odot}} \frac {\tau_{cz}} {\tau_{cz,\odot}} \right)^{-1},
\end{equation}
where $\omega$ is the angular velocity and $\tau_{cz}$ is the convective overturn timescale. 

Through the generation of magnetic fields, the stellar dynamo is linked to the X-ray emission in the Sun.
There is an empirical relationship between the X-ray luminosity, the bolometric luminosity, and Rossby number (\citealt{Wright2013}): 
\begin{equation}
\frac{L_X/L_{bol}}{L_{X,\odot}/L_{bol,\odot}} = Ro^{\beta} = \left( \frac{\omega}{\omega_{\odot}} \frac {\tau_{cz}} {\tau_{cz,\odot}} \right)^{-\beta}
\label{eqn:Lx}
\end{equation}
where the value of $\beta$, which describes the strength of the dynamo, has typically been assumed to be about -2, although some papers have argued for values closer to -3 (e.g. $\beta = -2.70\pm0.13$ from \citealt{Wright2013}). 
Like X-ray emission, mass loss occurs due to the magnetic field, which generates a hot, ionized stellar coronae.  
It is therefore not surprising that empirical relations between mass loss and X-ray flux have been found in the form of:
\begin{equation}
\frac{\dot{M}}{\dot{M}_\odot} = \left( \frac{F_X}{F_{X,\odot}} \right)^{\gamma},
\label{eqn:mdotfx}
\end{equation}
With values of $\gamma$ ranging from $0.66$ to $1.34 \pm 0.18$  \citep{Wood2005,See2017,Vidotto2021}.  

In general, these quantities are close to linearly proportional. We assume mass loss is linearly proportional to the X-ray luminosity in our models, i.e., the exponent on the right hand size of Eq.\ref{eqn:mdotfx} is equal to 1.

\begin{figure}
  \centering
	\includegraphics[width=\columnwidth]{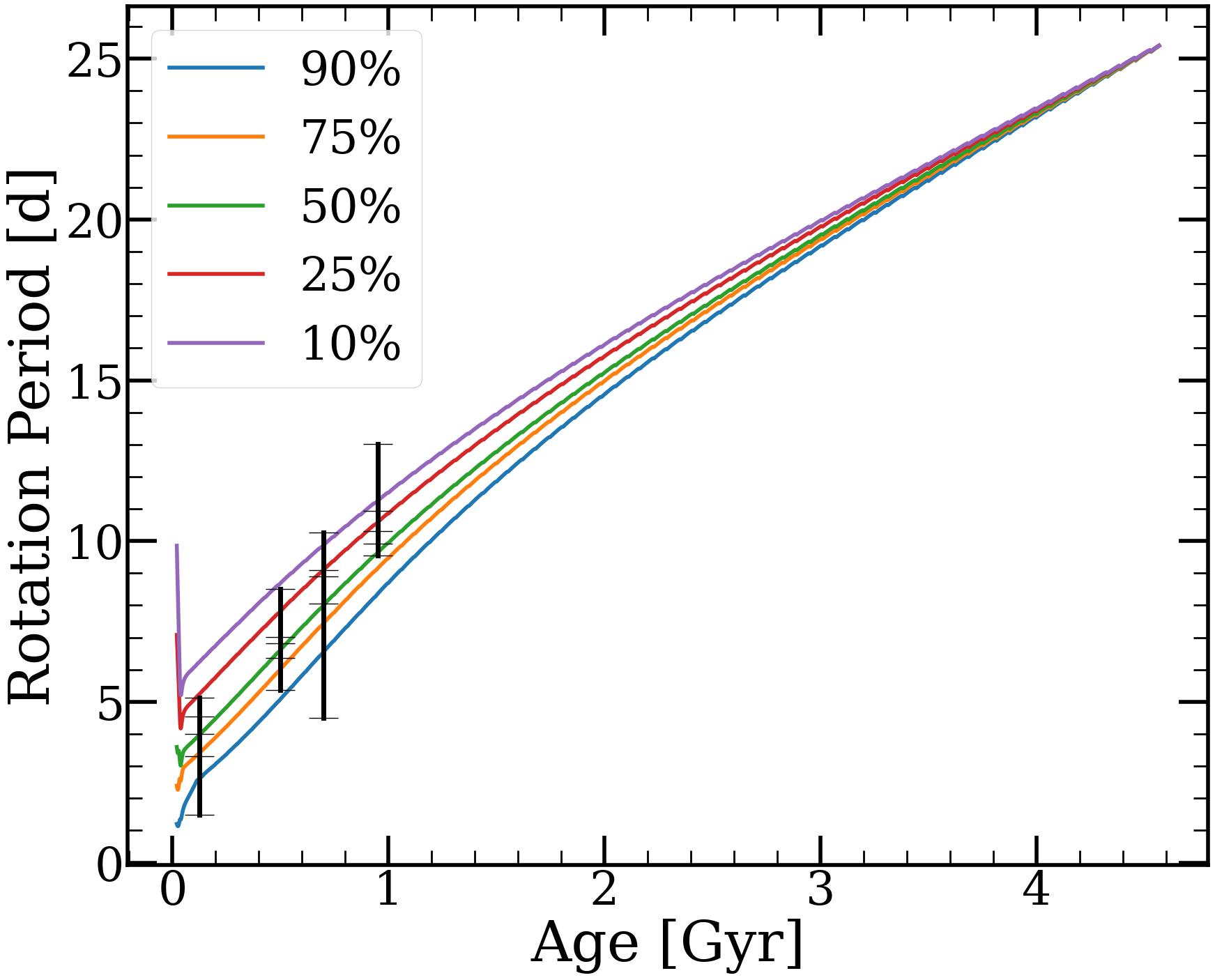}
    \caption{Empirical ranges of rotation rates in star clusters (points with ranges) compared with our model for the birth distribution of rotation rates. The rotation rates of our models are described by the legend, with the 90th (10th) percentile corresponding to the fastest (slowest) rotators. Observational data from several open clusters are plotted in black. From left to right, these are the Pleiades, M37, Praesepe, and NGC 6811. From top to bottom, the tick marks give the 10th, 25th, 50th, 75th, and 90th percentile rotation periods in these clusters. Cluster ages are adopted from \citet{GodoyRivera2021}.}
    \label{fig:Prot_grid}
\end{figure}

Combining the above equations and assuming the mass loss to be directly proportional to the X-ray luminosity, the mass loss rate is given by
\begin{equation}
\frac{\dot{M}}{\dot{M}_\odot} = \frac{L_{bol}}{L_{bol,\odot}} \left( \frac{\omega}{\omega_{\odot}} \frac {\tau_{cz}} {\tau_{cz,\odot}} \right)^{-\beta}
\end{equation}
Because $\beta$ describes the strength of the dynamo, it is also the parameter in our models which describes the slope of spindown versus time. 
We calibrate $\beta$ to the median rotation rates of 1 $M_\odot$ stars in the Pleiades, M37, Praesepe, and NGC 6811 reported in \cite{GodoyRivera2021} (see Fig.~\ref{fig:Prot_grid}). We expect our answer to lie in the range of $\beta$ = -2 to -3 (see e.g. the discussion in \citealt{Wright2013}). 

Torque occurs on the Sun due to mass loss from its magnetized stellar wind. From \cite{Matt2012}, the torque is related to the mass loss rate and dipole magnetic field strength by
\begin{equation}
\tau_W = K_1^2 B_*^{4m} \dot{M}_W^{1-2m} R_*^{4m+2} \frac{\Omega_*}{\left( K_2^2 v_{esc}^2 + \Omega_*^2 R_*^2 \right)^m}
\end{equation}
where $B_*$ is the magnetic field strength at the stellar equator, $\dot{M}_W$ is the total mass loss rate in the wind, $R_*$ is the stellar radius, $\Omega_*$ is the angular rotation rate of the stellar surface, $v_{esc}$ is the gravitational escape velocity, and $K_1 \sim 1.30$, $K_2 \sim 0.05$, and m $\sim$ 0.22 are dimensionless constants which were numerically determined through their simulations.
While we could have chosen from different forms to describe the torque, all modern prescriptions are tied to the same empirical constraints on the time history of rotation of solar analogs. As a result, they reduce to similar predicted time dependencies for mean field strength and mass loss rate.

Substituting the prescriptions for mass loss rate and the magnetic field, this results in the VP13 parameterization for angular momentum loss in a magnetized wind, which is expressed as: 
\begin{equation}
\label{eqndJdt}
\frac{dJ}{dt}\ =
  \begin{cases} 
      f_K K_M \omega \left(\frac{\omega_{crit}} {\omega_\odot} \right)^2 & \omega \frac {\tau_{cz}} {\tau_{cz,\odot}} \geq \omega_{crit} \\
      f_K K_M \omega \left(\frac{\omega \tau_{cz}} {\omega_\odot \tau_{cz,\odot}} \right)^2 & \omega \frac {\tau_{cz}} {\tau_{cz,\odot}} < \omega_{crit}
  \end{cases}
\end{equation}
The parameter $f_K$ is the solar calibration constant and is calibrated in our individual models to ensure that rotation rates are equal to the solar value at present day. The parameter $K_M$ is given by
\begin{equation}
\frac{K_M}{K_{M,\odot}} = c(\omega) \left( \frac{R}{R_\odot} \right)^{3.1} \left( \frac{M}{M_\odot} \right)^{-0.22} \left( \frac{L}{L_\odot} \right)^{0.56} \left(\frac{P_{phot}}{P_{phot,\odot}} \right)^{0.44}
\end{equation}
where $c(\omega)$ is the centrifugal correction term which is roughly 1. 

As seen in equation \ref{eqndJdt}, the angular momentum loss is saturated above some critical angular velocity ($\omega_{crit}$, the saturation threshold) scaled by the Rossby number, given by the ratio of the rotation period and the convective overturn timescale. This form is consistent with the behavior of a variety of diagnostics of stellar activity, such as X-ray emission, chromospheric activity indicators, and mean surface magnetic field strengths inferred directly or through star spot filling factors. All saturate in young, rapidly rotating stars. We adopt a saturation threshold of log $Ro = -0.677$ from \cite{Cao2022}, which was obtained by fitting to star-spot data. Above the saturation threshold $dJ/dt \propto \omega_{crit}^2 \omega$, while below this threshold $dJ/dt \propto \omega^3$. The effect of this is that initially fast rotating stars can remain fast rotating for longer, which further increases the mass loss observed in the highest mass loss models.

\subsection{Rotation}

The exact birth rotation of our Sun is unknown, so we consider a range of slow, moderate, and rapid rotators in our analysis. 
For clarity, slow, moderate, or rapid rotators roughly correspond to the 25th, 50th, or 90th percentiles of observed rotation rates, respectively. 
The youngest observed protostars have massive accretion disks that can exchange angular momentum with the star. Modern angular momentum calculations bypass the complexities of star-disk coupling by using empirical data in systems older than 10 Myr, where massive accretion disks are relatively rare.
In this study, we explored a set of birth conditions for the Sun with initial rotation periods corresponding to observed rotation periods of 0.9-1.1 M$_\odot$ stars in Upper Sco.
Models were initialized with a slower rotation rate at an earlier age and allowed to spin up to the desired rotation rate at 10 Myr. While it seems likely that the Sun was a slow or moderate rotator (e.g., only 15\% of solar mass stars in the Pleiades are rapidly rotating), 
we cannot rule out the possibility that the Sun could have been a rapid rotator, which would maximize its early mass and luminosity. We aim to address the question of whether any plausible solar history would be sufficient to explain the FYS problem. 

We explore both rotation as a solid body and rotation with partial core envelope coupling, also allowing the timescale for coupling to vary. 
See \citealt{Somers2016} for a detailed description of the models.
The mean angular velocity in the surface convection zone is only weakly dependent on depth. We therefore assume solid body rotation in convective regions at all times.

The radiative interior is more complex. From helioseismic data, the solar core is rotating nearly uniformly below a shear layer at the base of the solar convection zone. One limiting case model for internal transport of angular momentum is to assume solid body rotation in radiative regions; this would occur, for example, with strong magnetic coupling. However, there are other models where the spindown timescale is longer - for example, time scales of 10s of Myr are expected from wave-driven angular momentum transport. The spindown of stars is in better agreement with models including this "core-envelope decoupling", and the required timescales for the Sun are consistent with the observed flat rotation curve in the interior (\citealt{Denissenkov2010}; \citealt{Eggenberger2019}; \citealt{Somers2016}).
The results shown in our figures use rotational evolution with core envelope coupling. Our solid body models fail to properly reproduce the observed rotation periods of solar mass stars in young open clusters and thus we did not consider them further.

The parameters in our rotational models have been chosen to reproduce the spindown of the open clusters (Fig.~\ref{fig:Prot_grid}). The open cluster data is represented by the black lines. The $\beta$ value is calibrated to the median rotators (given by the green line) and $f_K$ is calibrated for each individual model. The starting point of these models is based on rotation data for 0.9-1.1 M$_\odot$ stars from Upper Sco (\citealt{Rebull2018}) with a median rotation period of 4.627 days (and a range of 14.33-1.52d for 10th-90th percentiles) at an age of 10 Myr, and all models have the solar rotation period of 25.4 days at 4.568 Gyr.
The best fit value for $\beta = -2.29$. 
As specified in the previous section, log Ro = -0.677 corresponding to $\omega_{crit} = 9.79 \omega_\odot$ is adopted from \cite{Cao2022}.

\begin{figure}
  \centering
	\includegraphics[width=\columnwidth]{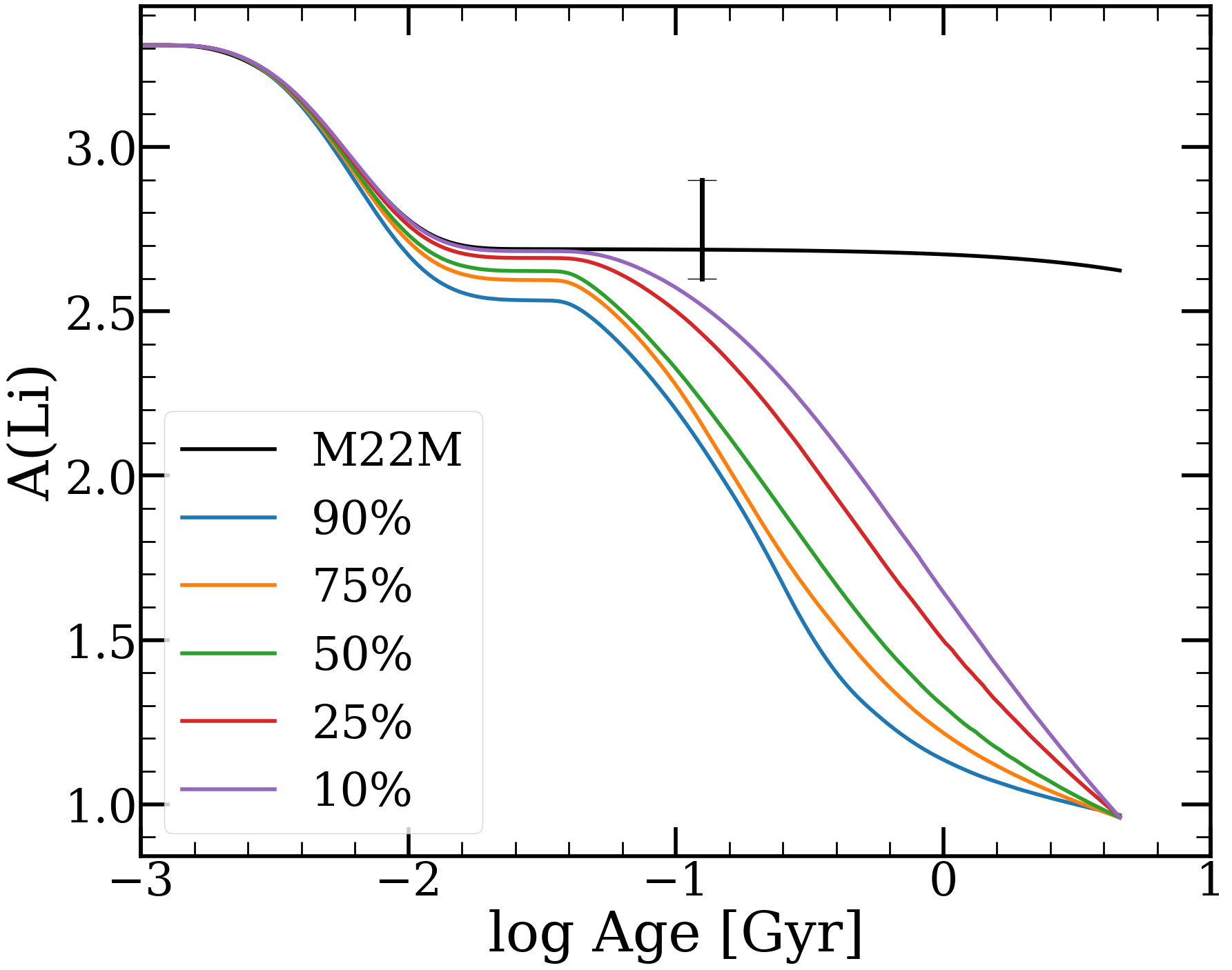}
    \caption{Lithium abundances over time in our reference model and rotational models from 10th-90th percentile of rotation rates. The values in the rotating models are calibrated to A(Li)=0.96 (\citealt{Asplund2021}). The black bar shows A(Li)=2.6-2.9 corresponding to the Pleiades at 125 Myr. In the non-rotating reference model, the present-day surface lithium abundance is too high.}
    \label{fig:lithium}
\end{figure}

The Sun is highly depleted in lithium compared to meteorites (\citealt{Lodders2021}). We calibrate the surface lithium abundance to the solar log relative abundance of A(Li) = 0.96 (\citealt{Asplund2021}) on the scale where the abundance of hydrogen is 12 via a mixing efficiency parameter. Holding this parameter fixed instead would have resulted in the lowest surface abundance of lithium in the rapid rotating case, and the highest surface lithium abundance in the slowest rotating case, so a range of solar rotation rates in practice translates to a range of efficiencies (see the discussion in \citealt{Somers2016}).
The mixing efficiencies $f_C$ and solar calibration constants $f_K$ for each model are shown in Table~\ref{tbl:rotcal}.
The lithium abundances are shown in Fig.~\ref{fig:lithium}. Some lithium is depleted in the early stages of solar evolution, prior to the main sequence. However, solar analogs in the Pleiades have lithium of about 2.6-2.9 at 125 Myr (see Fig 3 of \citealt{Bouvier2018}), implying 1.6 dex, or a factor of 40, depletion on the main sequence. Standard solar models predict little or no main sequence depletion, and the most promising explanation is rotationally induced mixing (\citealt{Pinsonneault1997}). There is a drastic difference between the non-rotating and rotating models that highlights the need for additional mixing processes to reproduce the present solar lithium abundance. Our reference and slowest rotating models show lithium depletion close to those seen in the Pleiades at the corresponding age, while our fastest rotating models are deficient in lithium. 

\begin{table}
\centering
\caption{Rotation Calibration Parameters}
\begin{tabular}{l l l}
\multicolumn{1}{c}{Rotation}
&\multicolumn{1}{c}{$f_K$}
&\multicolumn{1}{c}{$f_C$}\\
\hline
90\% & 7.7486 & 0.113 \\
75\% & 7.4566 & 0.274 \\
50\% & 7.2857 & 0.469 \\
25\% & 6.9192 & 1.13 \\
10\% & 6.6122 & 1.86 \\
\end{tabular}
\label{tbl:rotcal}
\end{table}
With these calibrations in mind, we move towards computing standard solar models which adopt parameters that are chosen to reproduce the behavior of solar analogs. Our new standard solar models adopt much of the same physics as our reference model, with a few key exceptions. Our standard solar models once again use YREC and this time include a full, self-consistent treatment of rotation, magnetism, and mass loss based on the physics presented in \S3. Now that we consider rotation histories in our YREC models, we must compute a suite of models from slow to rapid rotators to capture all plausible solar histories.

\begin{figure}
  \centering
	\includegraphics[width=\columnwidth]{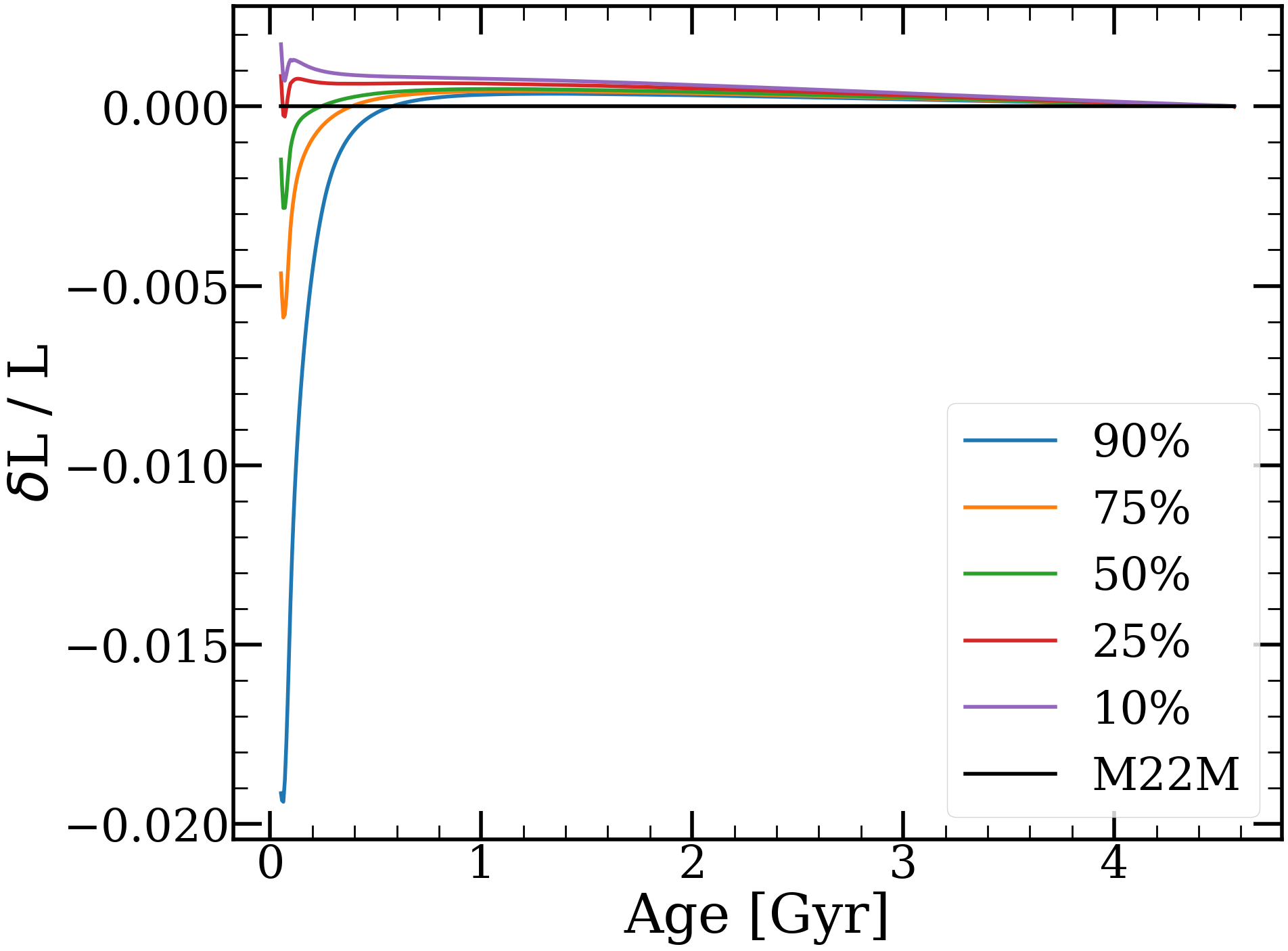}
    \caption{Fractional difference luminosities for rotating models compared to the reference M22M standard solar model. Differences are due to rotation only as mass is held constant at 1 M$_\odot$. The colors and percentiles of rotation rates are given in the legend, with the 90th (10th) percentile corresponding to the fastest (slowest) rotators. Rotation was initialized at a slower rate early in the pre-main sequence and allowed to spin up to the desired rate at 10 Myr.}
    \label{fig:rot_lum}
\end{figure}

The total angular momentum of the Sun is set by a combination of an assumed birth rotation rate and a disk coupling timescale; we can therefore achieve any given target rotation rate with a combination of the two. We adopt a fixed disk lifetime (0.1 Myr) and a range of birth rotation rates to reproduce the observed range in Upper Sco, our chosen system for initial conditions. We do so to avoid numerical problems in our most rapidly rotating cases. Our results for a range of birth rates and for a range of disk lifetimes are similar at later ages, so this choice does not affect our conclusions.

Fig.~\ref{fig:rot_lum} shows the fractional difference luminosities for the 10th-90th percentile rotators with a constant mass of 1$M_\odot$. Through much of the Sun's history, the luminosity of the Sun is consistent with non-rotating models. The slowest rotation periods (10th percentile) have the greatest change in luminosity at $\sim$100 Myr at roughly a tenth of a percent. 
There are two key competing effects which dominate at different times. First, rotation reduces the efficacy of diffusion on the main sequence, effectively making the star behave as if it had a lower metallicity and resulting in a hotter, more luminous star. Second, the stars are rotating relatively rapidly on the pre-main sequence and the centrifugal force from rotation partially counteracts the star's gravity. The star behaves like a slightly lower mass star, maintaining hydrostatic equilibrium at lower core temperatures, and is less luminous.

\subsection{Mass Loss}

\begin{figure}
  \centering
	\includegraphics[width=\columnwidth]{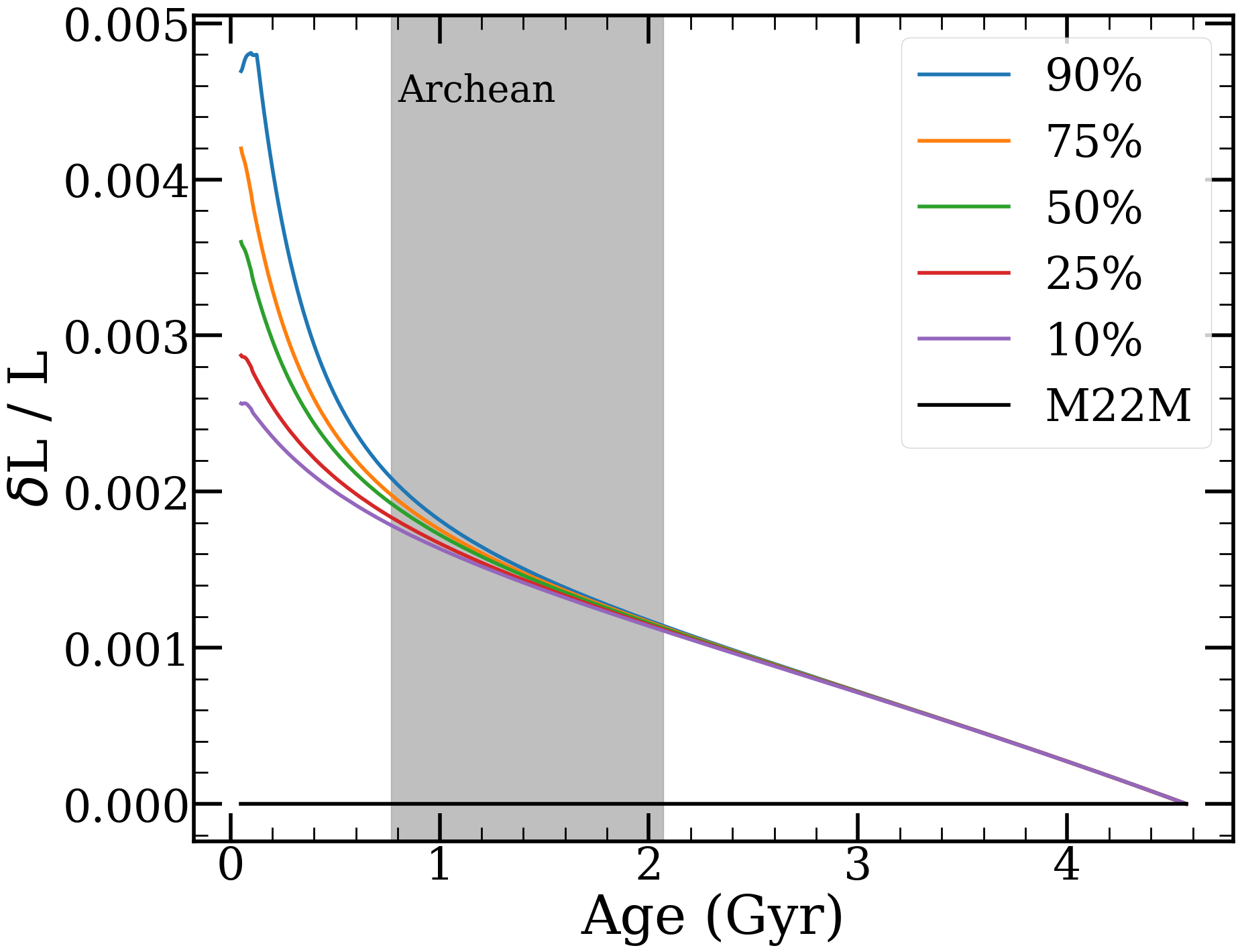}
    \caption{Difference plot for luminosities interpolated between two tracks with masses 1.00 $M_\odot$ (black line) and 1.00135 $M_\odot$ (not shown). The interpolation is based on the mass of the Sun as a function of time as computed in a grid of rotational models. 1.00135 $M_\odot$ was chosen to add the maximum mass loss found in our models to the mass of the Sun. In addition, the mass of the Sun as a function of time takes into account mass loss due to the Sun's luminosity. The majority of the mass loss occurs prior to 1 Gyr in fast-rotating models.}
    \label{fig:difference_RSCLM}
\end{figure}

We investigate the effect of mass loss on the Sun's luminosity using a tracer calculation. Determining the mass of the Sun as a function of time from our rotational evolution calculation allows us to interpolate the luminosity as a function of time from stellar models. We construct two non-rotating stellar models, the first with a mass of 1$M_\odot$ and the second with a mass of $1 M_\odot + {\delta}M$, where ${\delta}M = 1.35 \times 10^{-3} M_\odot$ is the maximum total mass loss in any of the rotational models. 
Using the mass as a function of time to compute the linear interpolation of the two models' luminosity tracks, the resulting luminosity as a function of age is shown in Fig.~\ref{fig:difference_RSCLM}. The colors in this plot each show a luminosity history corresponding to a different rotational model. Models that were initially faster rotating lost the most mass and show a greater increase in luminosity. 
Additional mass loss occurs as fusion converts some mass to energy, $m = E / c^2$, which is lost via the Sun's luminosity. Due to a technical detail of YREC, it is difficult to incorporate mass loss occurring in the core of the Sun into the full stellar models, but it is included in our tracer calculation in Fig.~\ref{fig:difference_RSCLM}. 

We observe an increase in luminosity of less than 0.5\% for the highest mass loss (fastest rotating) models. This is comparable to the difference between the AAG21 and M22M luminosities in Fig.~\ref{fig:l_diff_all} at very early ages and less than that during the Archean. Moreover, even in the lowest mass loss (slowest rotating) models, the fractional change in luminosity is comparable to what is observed in Fig.~\ref{fig:lum_errors}, demonstrating that realistic mass loss has a similar impact on the Sun's early luminosity to uncertainties in the input physics and a similar or smaller impact than uncertainties in the solar mixture. 
There are opposite trends in Figs.~\ref{fig:rot_lum} and~\ref{fig:difference_RSCLM}. Because the slowest rotator has the highest luminosity in the rotation-only case, it counteracts that the slowest rotator has the lowest luminosity in the mass-loss-only case. Effects of rotation dominate and the slowest rotator is the most luminous as a result. 
All cases including rotation and mass loss are more luminous than the reference model during the Archean.

\cite{Johnstone2015} also used rotational evolution models to estimate the mass loss in the stellar wind of the Sun. They used solid body models and observations of stellar rotation rates in young clusters to fit the mass loss rate as a function of radius, rotation rate, and mass. They found a dependence of mass loss $\dot{M}\propto \omega^{1.33}$ compared to our effective $\dot{M}\propto \omega^{2.29}$.
The early mass loss rate from our calculation is roughly 3-5x higher at 300 Myr (see Fig. 10 of \citealt{Johnstone2015}) and even greater at earlier times.
Their integrated mass losses are less than $3 \times 10^{-4} M_\odot$ even in their fastest rotating models, 4x smaller than we find in this paper. While they integrated from 100 Myr to 5 Gyr, which is a different timeframe than our work, assuming a reasonable saturation rate of mass loss for the first 100 Myr still leaves their total mass loss well short of our result. It is interesting to note that because of the feedback between mass loss and angular momentum loss, we arrived at estimates of total mass loss within a factor of three, despite our having adopted much higher mass loss rates for rapid rotators. 

Another notable difference is that they adopted solid body models, while we adopted core-envelope decoupling. Core-envelope decoupling matters less after the first $\sim$100 Myr because the evolution of the star's rotation becomes long compared to the coupling timescale ($\sim$20 Myr). This explains why they found good agreement between their observations and solid body models, while our solid body models did not reproduce the cluster rotation data from \cite{GodoyRivera2021}. However, both our calculation and theirs set strict limits on mass loss that are far less than those required to substantively impact the FYS problem.

\subsection{X-rays}

\begin{figure}
  \centering
	\includegraphics[width=\columnwidth]{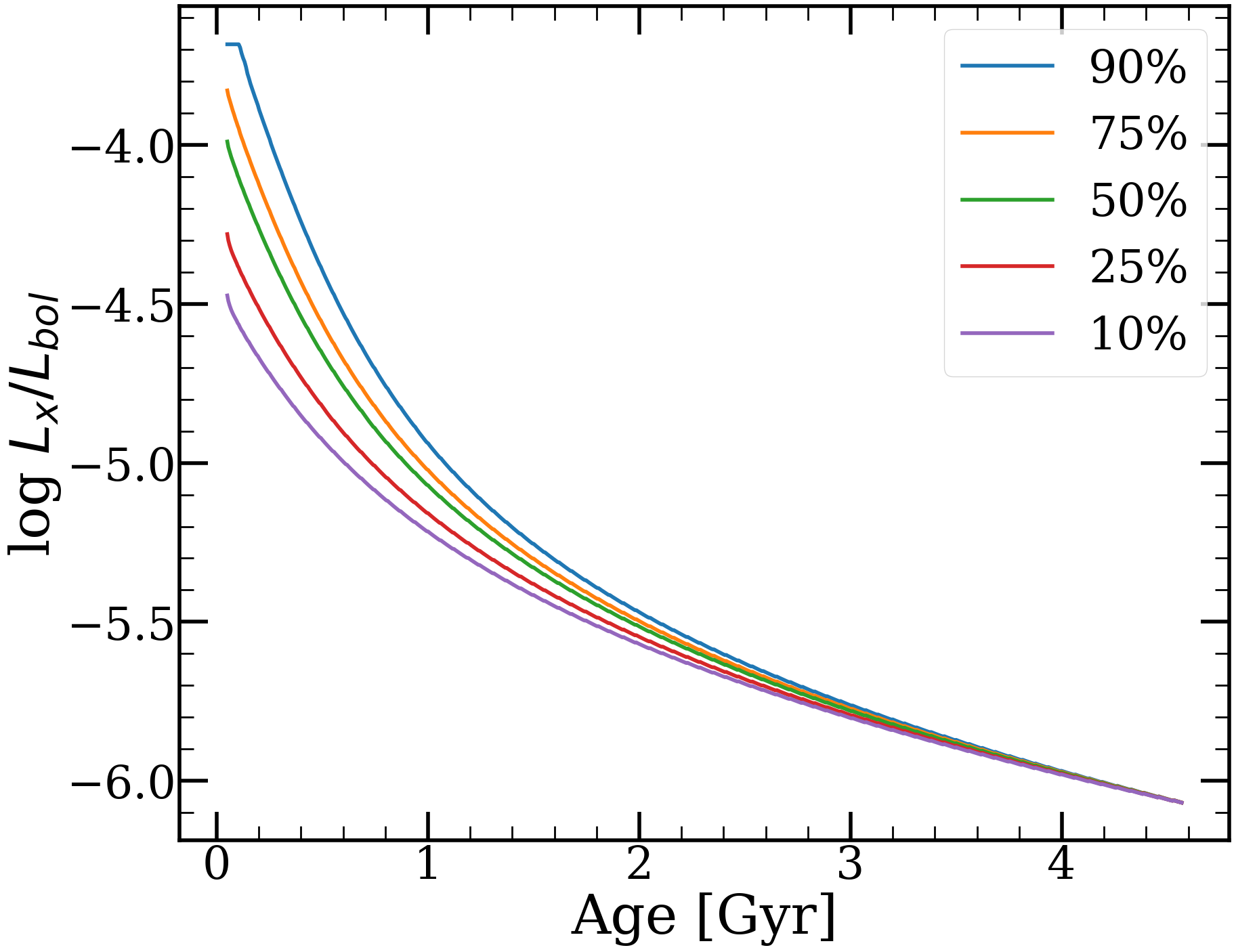}
    \caption{The ratio of the X-ray to bolometric luminosities from the rotational evolution models. The fastest rotators (90th percentile, blue) have the highest ratio of X-rays, with about an order of magnitude in spread to the slowest rotators (10th percentile, purple). The change in the X-ray emission calculated here is dominated by the angular velocity of the envelope, $\omega$ in equation \ref{eqn:Lx}.}
    \label{fig:Lx}
\end{figure}

The X-ray luminosities of solar type stars as a function of age have been measured (e.g. \citealt{Pizzolato2003}). The model that we adopt is consistent with the trends seen in the data, with higher X-ray luminosities and the saturation of X-rays among the most rapidly rotating stars. In addition to reporting other properties, we can therefore use our models to trace the energetic photon environment for the Sun in the past, and we use $L_x/L_{bol}$ as a convenient proxy.
The X-ray luminosity of the Sun (and in particular the high energy photons produced) could have a substantial impact on the atmospheres of the early planets through atmospheric escape or photodissociation, changing the early atmospheric chemistry of the planets. 

In our rotational evolution models, the X-ray luminosity is calculated using equation \ref{eqn:Lx}. 
The chosen solar parameters are $L_{X,\odot} = 10^{27.6}$ ergs s$^{-1}$, $\omega_\odot = 25.4$d, and $\tau_{cz,\odot}=12.3$d.
There is uncertainty in the choice of $L_{X,\odot}$ and $\omega_\odot$, while $\tau_{cz,\odot}$ is a parameter computed via models. The X-ray luminosity of the Sun varies over the solar cycle between $L_{X,\odot} = 10^{26.8}$ and $L_{X,\odot} = 10^{27.9}$ (\citealt{Judge2003}). While several choices for the solar X-ray luminosity could be adopted, we chose the arithmetic mean of these two values.
The rotation period of the Sun varies based on latitude from 24.5 days at the equator to 33.4 days near the poles. We chose the Carrington rotation rate of 25.4d, which is the rotation period observed for low latitude sunspots. 
Fig.~\ref{fig:Lx} shows the ratio of the X-ray to bolometric luminosities in our models. The proportion of X-rays is highest in the fastest rotating models with about an order of magnitude in spread, indicating substantial uncertainty for the young Sun based on its rotation history. 

\cite{Tu2015} used a rotational evolution model to calculate the X-ray emission as a function of time for solar-like stars. They adopt a range of possible rotation rates based on observations of h Per and NGC 6530, a slightly lower Ro = 0.13 (log Ro = -0.89) as a saturation threshold,
and a present day $L_X = 10^{27.2}$ ergs s$^{-1}$.
Their X-ray luminosity histories are computed using methods that are similar to ours and are consistent with our estimates.

\begin{table*}
\centering
\caption{Solar neutrino fluxes. Where not specified, units are  cm$^{-2}$ s$^{-1}$.}
\begin{tabular}{c c c c c c c c c c c}
\multicolumn{1}{c}{}
&\multicolumn{1}{c}{pp}
&\multicolumn{1}{c}{pep}
&\multicolumn{1}{c}{hep}
&\multicolumn{1}{c}{$^7$Be}
&\multicolumn{1}{c}{$^8$B}
&\multicolumn{1}{c}{$^{13}$N}
&\multicolumn{1}{c}{$^{15}$O}
&\multicolumn{1}{c}{$^{17}$F}
&\multicolumn{1}{c}{Cl}
&\multicolumn{1}{c}{Ga}\\
&\multicolumn{1}{c}{$\times 10^{10}$}
&\multicolumn{1}{c}{$\times 10^{8}$}
&\multicolumn{1}{c}{$\times 10^{3}$}
&\multicolumn{1}{c}{$\times 10^{9}$}
&\multicolumn{1}{c}{$\times 10^{6}$}
&\multicolumn{1}{c}{$\times 10^{8}$}
&\multicolumn{1}{c}{$\times 10^{8}$}
&\multicolumn{1}{c}{$\times 10^{6}$}
&\multicolumn{1}{c}{SNU}
&\multicolumn{1}{c}{SNU}\\
\hline
\textbf{Rotation, M22M} \\
90th & 5.97 & 1.44 & 7.97 & 4.93 & 5.46 & 3.21 & 2.50 & 5.27 & 7.90 & 126.90 \\
75th  & 5.97 & 1.44 & 7.98 & 4.91 & 5.41 & 3.18 & 2.46 & 5.19 & 7.83 & 126.60 \\
50th  & 5.97 & 1.44 & 7.99 & 4.90 & 5.37 & 3.15 & 2.44 & 5.14 & 7.78 & 126.40 \\
25th  & 5.98 & 1.44 & 8.01 & 4.87 & 5.30 & 3.10 & 2.39 & 5.04 & 7.69 & 126.00 \\
10th  & 5.98 & 1.44 & 8.01 & 4.85 & 5.25 & 3.08 & 2.37 & 4.98 & 7.63 & 125.70 \\
\hline
\textbf{No Rotation} \\
M22M & 5.96 & 1.43 & 7.91 & 5.04 & 5.72 & 3.44 & 2.68 & 5.66 & 8.23 & 128.50 \\
GS98 & 5.96 & 1.44 & 7.94 & 5.00 & 5.62 & 3.13 & 2.43 & 6.46 & 8.09 & 127.60 \\
AAG21 & 6.01 & 1.46 & 8.18 & 4.61 & 4.73 & 2.41 & 1.80 & 4.02 & 6.93 & 122.10 \\
M22P & 5.96 & 1.43 & 7.94 & 4.99 & 5.59 & 3.40 & 2.64 & 5.57 & 8.08 & 127.80 \\
\hline
\cite{Vinyoles2017} & 5.98 & 1.44 & 7.98 & 4.93 & 5.46 & 2.78 & 2.05 & 5.29 & &  \\
\cite{Serenelli2011} & 5.98 & 1.44 & 8.04 & 5.00 & 5.58 & 2.96 & 2.23 & 5.52 & &  \\
\cite{Bahcall2006} & 5.99 & 1.42 & 7.93 & 4.84 & 5.69 & 3.05 & 2.31 & 5.83 & 8.12 & 126.08 \\
\cite{Bahcall2001} & 5.96 & 1.40 & 9.3 & 4.82 & 5.15 & 5.56 & 4.88 & 5.73 & 7.7 & 129 \\

\end{tabular}
\label{tbl:neutrinos}
\end{table*}

\begin{table*}
\centering
\caption{Properties at the center of the Sun.}
\begin{tabular}{c c c c c c c c}
\multicolumn{1}{c}{}
&\multicolumn{1}{c}{$T_c$}
&\multicolumn{1}{c}{$\rho_c$}
&\multicolumn{1}{c}{$P_c$}
&\multicolumn{1}{c}{$Y_{init}$}
&\multicolumn{1}{c}{$Z_{init}$}
&\multicolumn{1}{c}{$Y_c$}
&\multicolumn{1}{c}{$Z_c$}\\
&\multicolumn{1}{c}{$10^{6}$ K}
&\multicolumn{1}{c}{g cm$^{-3}$}
&\multicolumn{1}{c}{$10^{17}$ erg cm$^{-3}$}
&\multicolumn{1}{c}{}
&\multicolumn{1}{c}{}
&\multicolumn{1}{c}{}
&\multicolumn{1}{c}{}\\
\hline
\textbf{Rotation, M22M} \\
90th & 15.582 & 148.8 & 2.322 & 0.2727 & 0.0180 & 0.6180 & 0.0188  \\
75th  & 15.575 & 148.7 & 2.322 & 0.2722 & 0.0179 & 0.6176 & 0.0187  \\
50th  & 15.569 & 148.7 & 2.321 & 0.2718 & 0.0178 & 0.6172 & 0.0186  \\
25th  & 15.558 & 148.5 & 2.319 & 0.2712 & 0.0177 & 0.6164 & 0.0185  \\
10th  & 15.550 & 148.4 & 2.318 & 0.2709 & 0.0176 & 0.6159 & 0.0184  \\
\hline
\textbf{No Rotation} \\
M22M & 15.622 & 149.3 & 2.327 & 0.2751 & 0.0185 & 0.6217 & 0.0194  \\
GS98 & 15.606 & 149.2 & 2.326 & 0.2735 & 0.0189 & 0.6196 & 0.0198  \\
AAG21 & 15.463 & 147.7 & 2.315 & 0.2641 & 0.0156 & 0.6097 & 0.0163  \\
M22P & 15.604 & 149.2 & 2.325 & 0.2737 & 0.0184 & 0.6202 & 0.0193  \\
\hline 
\cite{Vinyoles2017} &  &  & & 0.2718 & 0.0187 & 0.6328 & 0.0200 \\
\cite{Serenelli2011} & 15.62 & 151.4 & & 0.2724 & 0.0187 & 0.6333 & 0.0200 \\
\cite{Bahcall2006} & 15.67 & 152.9 & 2.357 & 0.2725 & 0.01884 & 0.6337 & 0.0202 \\
\cite{Bahcall2001} & 15.696 & 152.7 & 2.342 & 0.2735 & 0.0188 & 0.6405 & 0.0198 \\
\end{tabular}
\label{tbl:cent_prop}
\end{table*}

\begin{table*}
\centering
\caption{Surface elemental abundances and properties at the base of the convective zone.}
\begin{tabular}{c c c c c c c c c c c c c}
\multicolumn{1}{c}{}
&\multicolumn{1}{c}{$Y_s$}
&\multicolumn{1}{c}{$Z_s$}
&\multicolumn{1}{c}{$^7Li_s$}
&\multicolumn{1}{c}{$\alpha$}
&\multicolumn{1}{c}{R(CZ)}
&\multicolumn{1}{c}{$<\delta c / c>$}
&\multicolumn{1}{c}{M(CZ)}
&\multicolumn{1}{c}{T(CZ)}\\
&\multicolumn{1}{c}{}
&\multicolumn{1}{c}{}
&\multicolumn{1}{c}{}
&\multicolumn{1}{c}{}
&\multicolumn{1}{c}{R$_\odot$}
&\multicolumn{1}{c}{}
&\multicolumn{1}{c}{M$_\odot$}
&\multicolumn{1}{c}{$10^{6}$ K}\\
\hline
\textbf{Rotation, M22M} \\
90th  & 0.2477 & 0.01662 & 0.97 & 1.92 & 0.7145 & 0.00105 & 0.02414 & 2.18 \\
75th  & 0.2483 & 0.01661 & 0.96 & 1.92 & 0.7148 & 0.00105 & 0.02406 & 2.18 \\
50th  & 0.2488 & 0.01660 & 0.96 & 1.91 & 0.7150 & 0.00106 & 0.02399 & 2.18 \\
25th  & 0.2497 & 0.01658 & 0.96 & 1.91 & 0.7155 & 0.00108 & 0.02386 & 2.17 \\
10th  & 0.2503 & 0.01657 & 0.96 & 1.91 & 0.7158 & 0.00111 & 0.02378 & 2.17 \\
\hline
\textbf{No Rotation} \\
M22M & 0.2458 & 0.01667 & 2.62 & 1.93 & 0.7132 & 0.00112 & 0.02456 & 2.19 \\
GS98 & 0.2444 & 0.01706 & 2.65 & 1.94 & 0.7125 & 0.00111 & 0.02464 & 2.20 \\
AAG21 & 0.2350 & 0.01404 & 2.81 & 1.92 & 0.7197 & 0.00232 & 0.02246 & 2.10 \\
M22P & 0.2445 & 0.01663 & 2.65 & 1.93 & 0.7132 & 0.00112 & 0.02448 & 2.19 \\
\hline
\cite{Vinyoles2017} & 0.2426 & 0.0170 & & 2.18 & 0.7116 & 0.0005 & & \\
\cite{Serenelli2011} & 0.2429 & 0.0170 & 2.33 & 2.161 & 0.7124 & 0.0009 & & \\
\cite{Bahcall2006} & 0.2426 & 0.01697 & & 2.2097 & 0.7132 & 0.00099 & 0.02403 & 2.184 \\
\cite{Bahcall2001} & 0.2437 & 0.01694 & & 2.04 & 0.7140 & 0.00104 & 0.02415 & 2.18 \\
\end{tabular}
\label{tbl:conv_prop}
\end{table*}

\section{Implications for climate modelling and paleotemperature reconstructions}
\begin{figure}
  \centering
	\includegraphics[width=\columnwidth]{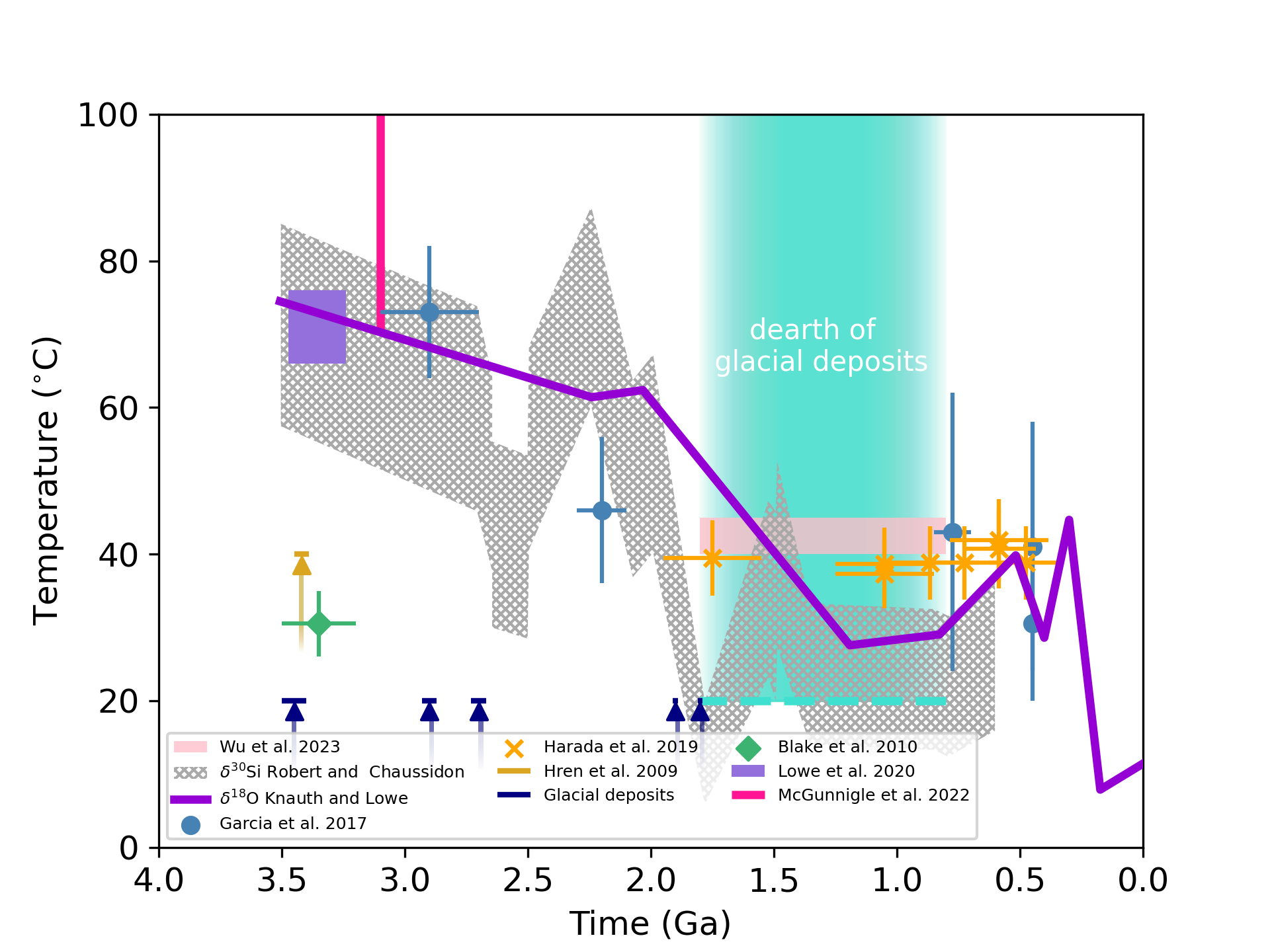}
    \caption{Archean and Proterozoic paleotemperature reconstructions derived from geolocical \citep{deWitt2016,Young1998,Ojakangas2014,Kuipers2013,Williams2005}, geochemical \citep{KNAUTH1978,Hren2009,Blake2010,Lowe2020,mcgunnigle2022triple} and biochemical \citep{Garcia2017,HARADA2019} proxies and the the evolution of Earth's spin \citep{Wu2023}. A lack of glacial evidence between 1.8-0.8 Ga implies temperatures $>$ 20 $^\circ$C.}
    \label{fig:temperature_recon}
\end{figure}

Earth's temperature evolution throughout the Archean and Proterozoic remains a topic of debate in the literature. A compilation of paleotemperature reconstructions is given in Fig.~\ref{fig:temperature_recon}.

Geochemical temperature reconstructions based on $\delta^{18}$O \citep{KNAUTH1976,KNAUTH1978} and $\delta^{30}$Si \citep{Robert2006} isotopic data from marine cherts have been interpreted as indicating temperatures > 30 $^{\circ}$C throughout most of the Proterozoic, even exceeding 60 $^{\circ}$C in the Archean and early Proterozoic. Various studies have questioned these result, citing e.g. the possibility of hydrothermal alterations, diagenetic processes, or challenging the central assumption that $\delta^{18}$O in seawater did not change over time (\citealt{vandenBoorn2007,Sengupta2020,Johnson2020}).
A newer analysis of cherts by \citet{Hren2009} uses $\delta^{18}$O in conjunction with hydrogen isotopes ($\delta$D) and estimates ocean temperatures $<$ 40$^{\circ}$C. Likewise, $\delta^{18}$O data derived from phosphates in sediments point to a temperate Archean with temperatures of 26-35 $^{\circ}$C at 3.2-3.5 Ga \cite{Blake2010}.

Despite the above criticisms and alternative analyses, interpretations of oxygen isotope data ($\delta^{18}$O, $\delta^{17}$O) as indicative of a very hot Archean $\sim$ 66-100 $^{\circ}$C have persisted in the literature \citep{Tartese2017,Lowe2020,mcgunnigle2022triple}.
These claims of moderate to high paleotemperatures based on isotope ratios are in conflict with evidence for glaciations occurring throughout the Archean and extending to the Paleoproterozoic \citep[e.g.][]{Young1998,Williams2005,Kuipers2013,Ojakangas2014,deWitt2016}. Later in Earth's history, a dearth of evidence for glacial deposits during the "boring billion" (ca. 1.8-0.8 Ga) suggests a prolonged period of elevated sea surface temperatures exceeding > 20 $^\circ$C. To explain why the modern length of day (lod) is 24 h, \citet{Wu2023} propose surface temperatures averaging 40-45 $^{\circ}$C during the boring billion, creating a resonance between the solar driven thermal tide and an atmospheric oscillation \citep{Zahnle1987a}. The resulting increased thermal torque would have effectively balanced the lunar tidal torque that would have otherwise lengthened the day to > 65 h, instead leading to a nearly constant lod for an extended time (lod lock).
However, geological evidence for a lod lock during the Precambrian is highly uncertain \citep{Laskar2024} and the tidal evolution of the Earth-Moon system may be explained without a resonant atmospheric tidal lock \citep{Farhat2022}. \citet{Farhat2024} find that the atmospheric resonance should have occurred later - in the Phanerozoic-  and that its amplitude is not high enough to cause a lod lock.

In recent years, a biochemical line of evidence has been invoked to bolster claims of gradual cooling of Earth's surface through time. Based on thermostability measurement of experimentally reconstructed ancestral enzymes, \citet{Garcia2017} report a decline in environmental temperature of photic zone organisms from $\sim$ 65–80 $^{\circ}$C in the Archean to $\sim$ 20-60 $^{\circ}$C in the Neoproterozoic. Similarly, \citet{HARADA2019} report environmental temperatures $\sim$40 $^{\circ}$C during the Proterozoic for marine phototrophic and planktonic organisms.
 
Some paleotemperature reconstructions from Fig. \ref{fig:temperature_recon} may be hard to reconcile with other constraints to Archean or Proterozoic Earth's properties proposed in the literature. According to nitrogen and argon isotope analysis of fluid inclusions in Archean hydrothermal quartz, the partial pressure
of N$_{2}$ was < 1.1 bar and may have been as low as 0.5 bar at 3.0-3.5 Ga \citep{Marty2013}. With the same method, $p$N$_2$ was constrained to similar or lower than modern at 3.3 Ga \citep{Avice2018}. Moreover, \citet{Som2012} limit air pressure at 2.7 Ga to less than
2 bar by studying fossilized raindrop imprints. The size distribution of gas bubbles in basaltic lava flows may even indicate an upper limit of half of modern air pressure at 2.7 Ga \citep{Som2016}. 

\begin{figure}
  \centering
	\includegraphics[width=\columnwidth]{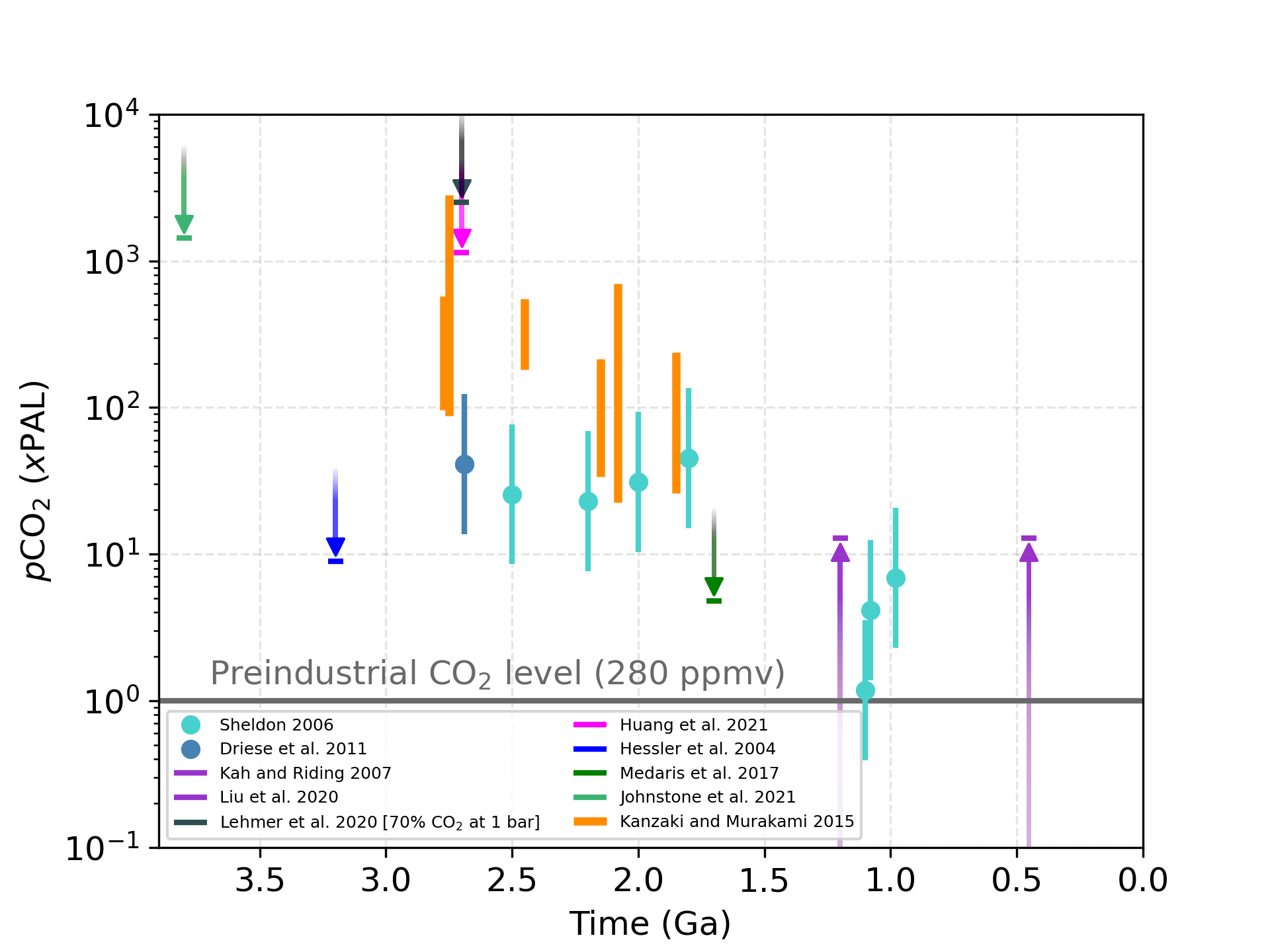}
    \caption{Reconstructions of Archean and Proterozoic atmospheric CO$_2$ relative to preindustrial atmospheric levels (PAL) of 280 ppm.}
    \label{fig:CO2_recon}
\end{figure}

Past CO$_2$ concentrations may have reached much higher values than the preindustrial atmospheric CO$_2$ level (PAL) of 280 ppm. Atmospheric escape modelling implies that the Archean atmosphere at 3.8 Ga may have required a minimum of 40\% CO$_2$ to prevent rapid atmospheric loss \citep{Johnstone2021}. \citet{Hessler2004} derive a minimum of $2.51\times10^{-3}$ bar CO$_2$ at 3.2 Ga by evaluating weathering rinds of river gravels. 2.7 Ga old fossil micrometeorites exhibiting oxidation \citep{tomkins2016} have spurred a number of studies trying to constrain $p$CO$_2$ with inferred lower abundance limits ranging from 23 - 70\% \citep{Payne2020,Lehmer2020, Huang2021}, and some authors additionally considering low $p$N$_2$ \citep{rimmer2019,Payne2020} as explanation. \citet{Sheldon2006} developed a weathering mass-balance approach to estimate $p$CO$_2$ from paleosols that was subsequently used in other studies \citep{Driese2011,Sheldon2013,Medaris2017}. Estimates from this method reach up to $\sim$ 40-45 PAL in the Archean and Paleoproterozoic. A different method by \citet{Kanzaki2015} relies on the cation concentrations in porewaters at the time of weathering of paleosols to estimate $p$CO$_2$ between 2.77-1.85 Ga. Their estimates imply a gradual decline in $p$CO$_2$ during the Neoarchean to Paleoproterozoic period and exceed the values calculated by \cite{Sheldon2006}'s method roughly up to one order of magnitude.
Consistent with the low CO$_2$ values found by \cite{Sheldon2013} for the late Mesoproterozoic, cyanobacteria microfossils finds exhibiting calcification - a process thought to only occur below the threshold of $\sim$ 13 PAL - further only permit moderately elevated CO$_2$ concentrations for 1.2 Ga \citep{Kah2007} and 0.4 Ga \citep{Liu2020}.

In Fig. \ref{fig:clima_inverse}, we present a set of simple 1-D climate model calculations (See Appendix \ref{app:clima} for a model description). The "inverse" climate model (originally by Kasting et al., used in e.g. \citealt{Kopparapu2013}) receives a parameterised temperature-pressure profile and a set of greenhouse gas concentrations as input and calculates the solar constant $S$ needed to sustain the given surface temperature. We calculate required solar constant values to reach various target surface temperatures for the CO$_2$ constraints sets from Fig. \ref{fig:CO2_recon} using a fixed pN$_2$ value of 1 bar. We additionally apply a conservative CH$_4$ amount of 20 mbar for all calculations. This value is consistent with the lower limit for Archean Sulfur mass-independent fractionation \citep{Zahnle2006} and roughly corresponds to predictions by biogeochemical boxmodelling coupled to photochemistry \citep{claire2006}.

\begin{figure}
  \centering
	\includegraphics[width=\columnwidth]{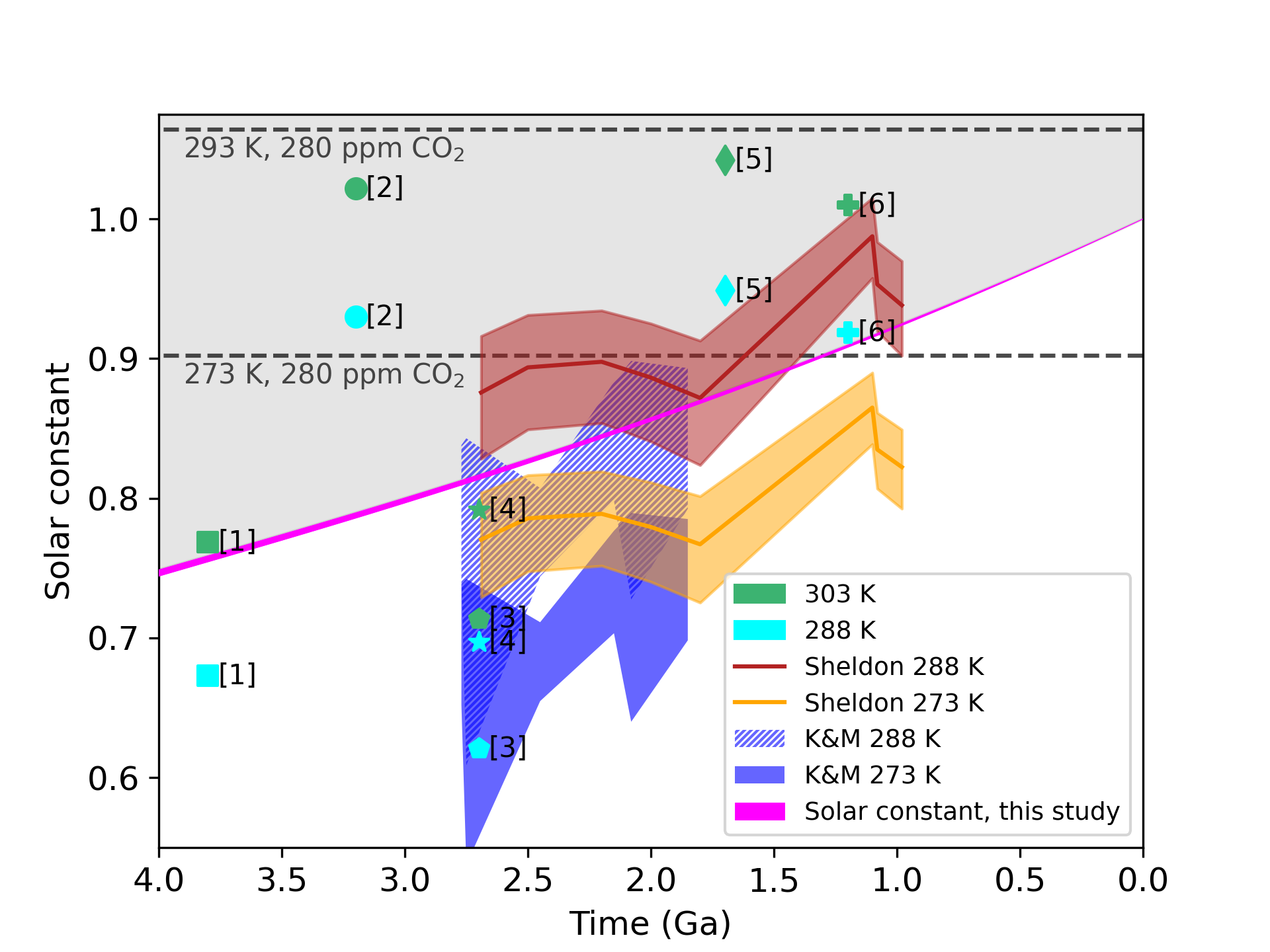}
    \caption{ 1-D "inverse" climate model calculations results. The grey shaded region indicates where the estimated solar constants required to maintain a given surface temperature exceeds the range predicted by the solar luminosity model (magenta). Calculation for CO$_2$ constraints based on Sheldon's method \citep{Sheldon2006,Driese2011} for 288 K and 273 K are shown in brown and orange, respectively. Calculation for the \citet{Kanzaki2015} CO$_2$ constraints are shown for 288 K (hatched blue) and 273 K (solid blue). Results for the CO$_2$ [1] lower limit by \citet{Johnstone2021}, [2] lower limit by \citet{Hessler2004}, [3] lower limit by \citet{Lehmer2020}, [4] lower limit by \citet{Huang2021}, [5] lower limit by \citet{Medaris2017} and [6] upper limit by \citet{Kah2007} are given for 303 K (green) and 288 K (cyan). Grey dashed lines are for a 1 bar N$_2$ atmosphere with 1 PAL CO$_2$ and no methane.}
    \label{fig:clima_inverse}
\end{figure}

Our results show that solar luminosity is high enough to maintain temperate conditions above freezing or similar to modern Earth's 288 K, even with our conservative assumptions on greenhouse warming by CH$_4$. However, we struggle to reconcile CO$_2$ constraints and conservative CH$_4$ assumptions with very high Archean temperatures or high Proterozoic temperatures. This is not unexpected: For example, \citet{Charnay2017}'s GCM calculations need 1 bar CO2 and 2 $\times 10^{-3}$ bar CH$_4$ with 1 bar N$_2$ to reach 68 $^{\circ}$C at 3.8 Ga. To obtain a 40-45 $^{\circ}$C temperature average over the boring billion, \citet{Wu2023} used a surface pressure of 2 bar in conjunction with 100-200 mbar of CO$_2$, vastly exceeding \citet{Sheldon2006}'s range ($\sim$ factor 500) and the upper limit set by \cite{Kah2007} ($\sim$ factor 40). 

Our 1-D climate calculations demonstrate a clear mismatch between the solar constant values needed to reach high temperatures > $\sim$30 $^{\circ}$C and the narrow range of possible solar luminosity determined in this study. In contrast, the difference in required luminosities implied by the errors given in CO$_2$ reconstructions alone e.g by \citet{Kanzaki2015} may reach 10\% $S$. Any attempt of reconciling very high temperature reconstructions, a low pressure atmosphere and literature CO$_2$ reconstructions via the minimal uncertainties in solar luminosity our modelling predicts seems unviable. That said, maintaining cooler-but-still clement conditions that would avoid global glaciation ($\sim$0 $^{\circ}$C < T < $\sim$30 $^{\circ}$C) seems plausible given the solar constants in this work, and our simple climate model simulations.

\section{The New Standard Solar Model}

In \S3, we demonstrated that the impact of non-standard physics on classical solar models like our reference model is generally a small perturbation over the time evolution of the models. Nonetheless, there are certain properties of the Sun that can only be reproduced via models which include a more complete understanding of physics. For example, we showed that our work agrees with prior results that rotational mixing is necessary to reproducing the lithium depletion in the Sun. In this section, we describe the results from our rotating YREC models, which includes rotation and mixing in a set of standard solar models for the first time. The inclusion of rotation and magnetism in our models also allows us to also include calculations of the X-ray luminosity as a proxy for solar activity, making the outputs of this set of standard solar models useful to a broader set of studies.  
Our results are expanded to a standard solar evolutionary history rather than simply a present-day standard solar model. 
Classical solar models match today's properties of the Sun, but our models are also constrained by relevant observations - rotation periods in particular. The slope of spindown has been calibrated to data from open clusters, the initial rotation rates of our models match stars from Upper Sco at 10 Myr, the saturation threshold is obtained via star-spot data, and the mixing efficiency is calibrated to match surface lithium.

We will highlight a median rotator as the best singular result for the Sun's evolutionary history, citing the range from the 10th to 90th percentiles of rotation rates as our uncertainty. Because it was used for the calibration, the median rotation in Fig.~\ref{fig:Prot_grid} most closely matches medians in cluster data (compared to slow or rapid rotators and their matching percentiles). It also naturally lies in the intermediate region between the slow and rapid rotators. We again caution that the Sun's past rotation history is uncertain and not well constrained, so the reader may wish to consider whether the full range of allowed rotation is relevant to their studies. 
Regardless of which initial rotation rate we choose, the overall picture of the Sun's evolution will be similar because stellar rotation quickly converges on the main sequence within a few 100 Myr. 

Tables \ref{tbl:neutrinos}, \ref{tbl:cent_prop}, and \ref{tbl:conv_prop} summarize the relevant results from all of our solar models and selected results from the literature. 
Results from YREC models adopting the M22M abundances and including rotation are grouped together, separated by a horizontal line. The values are the percentiles of rotation rates, with the 90th percentile corresponding to rapid rotators.
Results from non-rotating models adopting various metal abundances are grouped together, separated by another horizontal line from results from the literature. 

Table \ref{tbl:neutrinos} describes the neutrino fluxes in our models. Neutrino fluxes in this work are consistent with those found in the literature, with the biggest discrepancy generally being the non-rotating model adopting AAG21 metal abundances and mostly showing slightly lower (but still consistent) neutrino fluxes. Additionally, any rotation slightly decreases the neutrino fluxes for heavier elements since they do not settle as much.

Table \ref{tbl:cent_prop} describes the properties at the center of the Sun.  Our central helium abundance is lower than in the literature with a fractional change of $\sim 0.01-0.02$. Stellar interiors models use a Taylor series expansion in the deep core, and the precise central abundance depends on the inner fitting point. The central temperatures and density, by contrast, are integrated to the center and insensitive to the fitting point location. As a consequence, there is slight disagreement for the central abundances since they are reported from the inner fitting point.
 
Table \ref{tbl:conv_prop} describes the properties at the surface of the Sun and the base of the convective zone. The rotation and no rotation cases have generally good agreement with the surface helium ($Y_s \sim 0.248$) and the radius of the base of the convective zone (0.7132 R$_\odot$). The fractional changes in sound speed, $<\delta c / c>$, are similar in each case.
Only the cases with rotation reproduce the solar lithium abundance. When higher mixing efficiencies are needed to match the lithium (see Table \ref{tbl:rotcal}), it also increases the surface helium abundance. Helium will otherwise undergo settling in the Sun and a greater mixing efficiency returns more helium to the surface.

\begin{figure}
  \centering
	\includegraphics[width=\columnwidth]{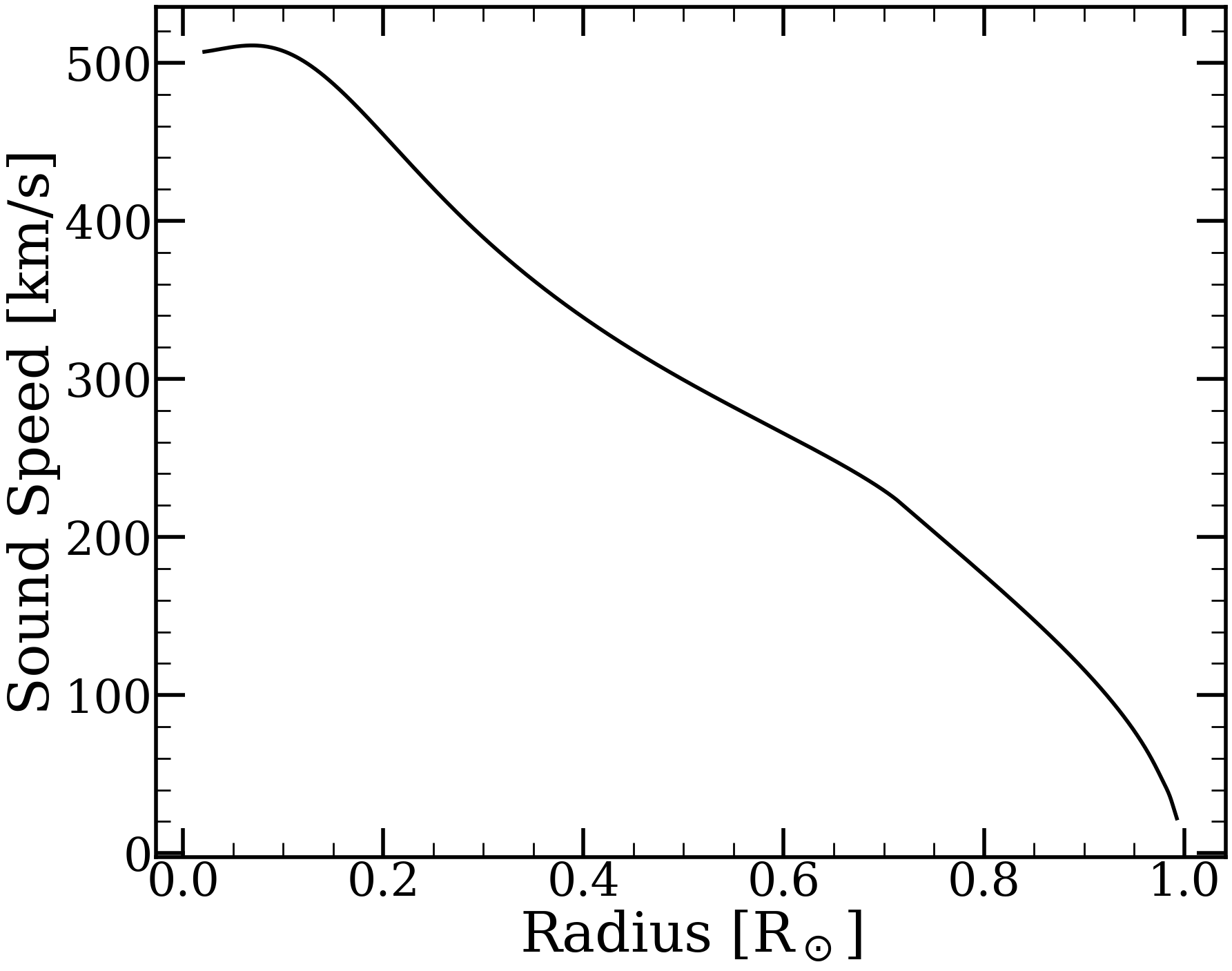}
    \caption{The present day sound speed as a function of radius as calculated in our models.}
    \label{fig:csound}
\end{figure}

Fig.~\ref{fig:csound} shows the present day sound speed profile of the Sun in our reference model. It peaks above 500 km/s in the core and decreases to less than 100 km/s in the outer envelope. The sound speed profiles can be used as one test to help identify which solar compositions best match solar observations, so the sound speed presented here serves as a baseline for reference in later plots.
Sound speed difference plots are measured relative to the observed sound speed profile of the Sun. Fig.~\ref{fig:csound} differs from the observed sound speed profile at roughly the tenth of a percent level. 

\begin{figure}
  \centering
	\includegraphics[width=\columnwidth]{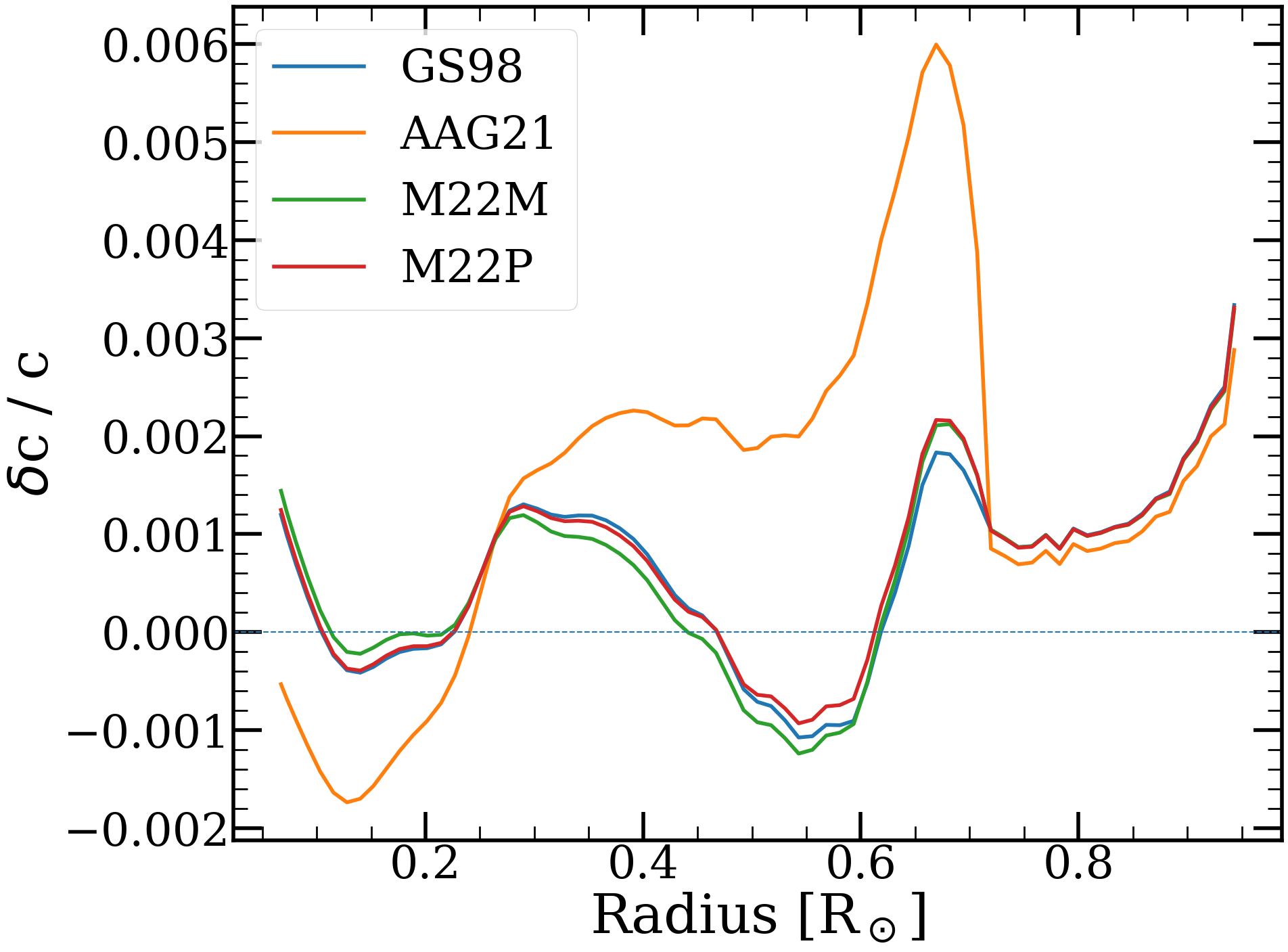}
    \caption{Fractional difference of the sound speed between models and observations. Each model uses a different opacity table corresponding to different elemental abundances - GS98 (blue), AAG21 (orange), M22M (green), and M22P (red). The fit of AAG21 is dramatically worse due to lower metal abundances, in particular the abundances of C, N, and O.}
    \label{fig:csound_all}
\end{figure}

\begin{figure}
  \centering
	\includegraphics[width=\columnwidth]{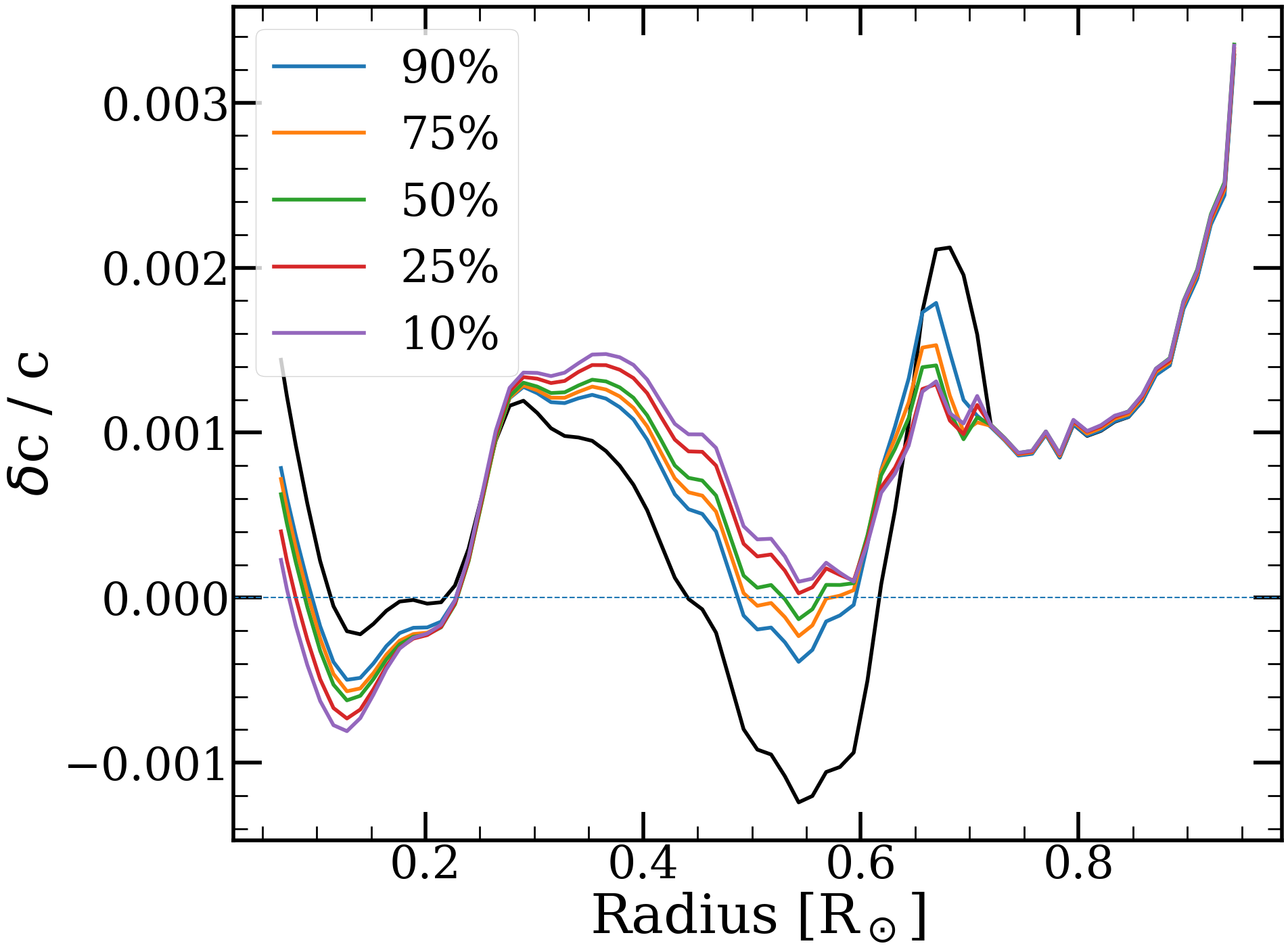}
    \caption{Fractional difference of the sound speed between models and observations. The black line is our non-rotating reference model (M22M) and the colored lines correspond to the different percentiles of rotation rates.}
    \label{fig:csound_rot}
\end{figure}

As in the literature, we observe a difference in the sound speed profiles of our models. Figs.~\ref{fig:csound_all} (no rotation) and~\ref{fig:csound_rot} (rotation) compare the sound speeds in the various models to the sound speed as presented in \cite{Basu2000}. The observed sound speed profile of the Sun is determined via helioseismic inversions using solar frequencies from the Michelson Doppler Imager (MDI) data set. A change of 0 is shown by the dotted line. In Fig.~\ref{fig:csound_all}, the agreement between our models and observations is best for GS98 and worst for AAG21, with the difference emphasized near the base of the convective zone at a radius R=0.7132 R$_\odot$ (also see $<\delta c / c>$, the root-mean-square of the fractional difference in sound speed, in Table~\ref{tbl:conv_prop}). This result is in good agreement with results presented in other works (e.g. \citealt{Magg2022}, others). 
The sound speed agreement favors higher metallicity compositions and factors into our selection of the M22M abundances for our standard solar models.

Fig.~\ref{fig:csound_rot} shows the effects of rotation on the sound speed profile. The rotating models show less severe peaks and valleys. Near the base of the convection zone, this occurs because rotational mixing smooths the effects of diffusion. If you change the helium content there, it must change elsewhere, and we see an increase in $\delta c /c$ in the deeper layers of the Sun. Changes in the sound speed profile are correlated due to mass conservation and the dependence of the sound speed on density ($c_s^2 = dP/d\rho$). In both the rotating and the non-rotating cases, the disagreement between the observed and modeled sound speeds increases near the surface of the Sun. This is due to the choice of radius to which our models are calibrated, specifically that we adopted the radius from \cite{IAUB3} (a smaller radius shows better agreement with the observed sound speed profile).

\section{Discussion}

The Faint Young Sun problem could be viewed as a problem for our understanding or the Sun, or a problem for our understanding of the Earth. When viewed through the lens of stellar evolution theory, the evolution of the Sun is driven by fundamental physical properties, and there are powerful reasons to expect solar evolution to be insensitive to changes in standard input physics. The solar calibration itself is an important consideration: solar models must reproduce the current solar properties, dramatically reducing the dynamic range of possible model trajectories. In addition, important current solar global properties - mass, age, radius and luminosity - are now exceptionally well measured. We confirm that the level that the uncertainties in solar modeling contributes to the solution is small. At most, perturbing the input physics to our models show a maximum variation in luminosity of 0.2\%. Recent controversies about the solar heavy element mixture modestly impact the luminosity at a given time by up to 0.3\%. Neither effect is substantive enough to impact the Faint Young Sun problem.

However, there have been models proposed where physical processes neglected in the standard framework, in particular mass loss, could lead to very different histories. We have therefore investigated mass loss in the context of a physically self consistent model that also incorporates rotation. We find that mass loss and rotation can only induce small changes, at the 0.3\% level in L. The underlying reason is that the current solar mass loss rate is negligible, so the only avenue for a high solar mass loss rate is to have the Sun experience much stronger winds in the past.  Wind strength is, indeed, higher in rapid rotators; however, most solar analogs rotate slowly for almost their entire main sequence lifetime, and virtually all rotate only $\sim 3$ times the solar rate at roughly fifteen percent of the solar age. Any self consistent solar wind model is thus constrained to minimal main sequence mass loss, severely limiting the impact on the history of the Sun. 
Independent calculations by Johnstone et al. 2015 found even smaller effects. Mass loss rates derived from full evolutionary models are therefore below empirical estimates (a best estimate of $6-7 \times 10^{-3} M_\odot$; \citealt{Wood2002}). 
Combining all of the effect that we identified these leads to an increase in luminosity at any given time at the percent level or less. We conclude that the Sun does not cause the Faint Young Sun problem, and the solution should be found closer to home. 

Fortunately, we can now build terrestrial climate models with a robust and well predicted solar luminosity evolution.
Incorporating paleotemperature reconstructions and predictions for greenhouse gas concentrations into climate models demonstrates that is reasonable for our solar luminosity history to reproduce temperate conditions on Earth. However, given the small uncertainties in luminosity, such models fall well short of reproducing the higher temperature reconstructions during the Archean. 

We find that the most substantial increase in luminosity in our models occurs before the Archean. The mass loss that occurs according to a prescription for mass loss via a magnetized wind primarily occurs early in the life of the Sun (see Fig.~\ref{fig:difference_RSCLM}). Additionally, the compositional uncertainties and input physics uncertainties show the potential for their greatest changes early in the Sun's history (Fig.~\ref{fig:l_diff_all} and ~\ref{fig:lum_errors}). The potential impact of this result is greater for Mars than it is for Earth. As Mars had conditions for liquid water at an even earlier age than Earth, the FYS problem is more severe for the red planet and contributions from any source may be significant. 

One avenue through which the Sun could lose additional mass that we did not explore in this paper is coronal mass ejections (CMEs). Mass loss due to CMEs may be greater than mass loss in the steady solar wind (see \citealt{Cranmer2017}). We caution that any mass loss due to CMEs will, however, result in angular momentum loss and spindown, lowering the mass loss rate due to the solar wind. There is a finite angular momentum for young stars and as such, a finite amount of angular momentum that can be lost to arrive at the present day Sun. The interplay between CMEs and angular momentum loss may make the total mass loss comparable to our models which only consider angular momentum loss due to a wind. 


Given that a changing mass for the Sun changes its orbital separation with the planets, there may be observable signatures on Earth or Mars that could reveal the Sun's early history if its mass loss is large enough. \cite{Spalding2018} found that measuring Milankovitch frequencies could test the Sun's mass to a precision of $\sim$1\% using a link between the banding of sediments and the orbital properties of Earth or Mars. We note that the $\delta$M we calculated is much less than the precision of such a measurement. Nonetheless, it would be useful to learn whether a more massive young Sun is observationally disfavored. 


The stellar spectrum is an important ingredient for the habitability of planets. In particular, the high-energy UV and X-rays dictates much about the atmospheres. The high-energy flux can heat and inflate the atmosphere, leading to increased rates of evaporation (e.g. \citealt{Wordsworth2013}). In addition, it impacts atmospheric chemistry, photodissociating molecules or driving chemical reactions, and groups have derived estimates of the early solar spectrum (e.g. \citealt{Claire2012}). Knowledge of the UV is crucial for the interpretation of possible biosignatures. The concentrations to which gases such as O$_2$ and CH$_4$ build up in an atmosphere are tied to their destruction rates through photochemistry as well as their production rates \citep{Meadows2010}.

Star spots could further complicate the stellar spectrum. The fraction of a star's surface that is covered by star spots is higher for rapidly rotating stars and the radius of these stars is also inflated. \cite{Cao2022} used a two-temperature model to describe the appearance of these stars. Studying the effect this has on stellar spectra could be an interesting direction for future study. We note that star spot filling factors are already low for solar mass stars in young open clusters, however, so the effect will likely be modest. Finally, a direct model including mass loss could capture non-linear effects that could be of interest in precise solar models.

\section{Conclusions}

The FYS problem is a longstanding problem in astronomy. Several solutions have been proposed. While the most likely solution is a combination of different effects that must balance surface temperatures, atmospheric compositions, etc. with the solar luminosity, this paper aims to address the extent to which uncertainties in solar modeling can contribute with a particular interest in the mass loss history of the Sun.

We use YREC stellar models and perturb the input physics to investigate the effect on the evolutionary tracks. The result is that the uncertainty is small - less than 0.2\% (0.382-0.299 W m$^{-2}$ from 3.8-2.5 Ga) even at the beginning of the Archean when the uncertainty is greatest - which is far less than the 21-33\% increase in luminosity (59-42 W m$^{-2}$) that would be required for solar modeling to solely solve the FYS problem. We also investigated the effects of composition, finding a slightly larger increase in luminosity of 0.3\% (0.596-0.421 W m$^{-2}$) for models adopting the AAG21 metal abundances. 

We then consider rotation and mass loss. Stars are born with a range of rotation rates, which are mostly slow even at young ages because of interactions between the star and its disk. The feedback between mass loss and angular momentum leads to spin down on the main sequence, with rotation rates converging even for relatively young open clusters (e.g. Praesepe, 700 Myr). 
For slow rotators, Skumanich implies $\dot{J} \propto \omega^3$, with wind loss rates scaling with rotation at comparable rates (see \S3). This saturates at a level around 10x solar, and the saturated domain corresponds to high mass loss.
However, the duration of the saturated phase is brief for solar analogs, and mass loss rates rapidly approach the low solar value once it has ended.

Previous standard solar models assumed a constant mass for the star over the course of its evolution. We find that including mass loss can increase the luminosity of the Sun up to 0.3\% (0.468-0.335 W m$^{-2}$). The magnitude of this increase is dependent on the birth conditions of the Sun's rotation, with rapid early rotation corresponding to greater mass loss and thus a higher luminosity. Combining this and the previous effects yields a total increase in luminosity less than 0.5\% (0.849-0.616 W m$^{-2}$).

We present compilations of paleotemperature reconstructions and atmospheric compositions for the Earth in \S4. We use inverse climate models to compute the solar constant $S$ required to produce temperate surface temperatures. We find that our solar luminosity and its uncertainty is consistent with a temperate Earth, but it would not easily reproduce higher temperatures. 
There is controversy over temperature reconstructions $>60^\circ$C, but separate from this controversy it is worth noting that temperature estimates are $>20^\circ$C through much of Earth's history. This indicates that the magnitude of the FYS problem is actually worse than simply reproducing a temperature comparable to the present day surface temperature of 288 K.

We present the luminosity, solar constant, X-ray luminosity, and the $T_{\rm eff}$ as a function of age for our solar models. The increase in these properties (other than $T_{\rm eff}$) at early ages could be interesting with regard to the atmospheres and habitability of young exoplanets as the increase in high energy photons could lead to more atmospheric evaporation or change the atmospheric chemistry. Future work could aim to include the effects of star spots on the stellar spectrum and whether that alters the impact of young host stars on their exoplanets.

\section*{Acknowledgements}

B.S.G. was supported by the Thomas Jefferson Chair for Space Exploration endowment from the Ohio State University.
S.D.D.G and S.T.B. acknowledge support from the GSFC Sellers Exoplanet Environments Collaboration (SEEC), which is supported by NASA’s Planetary, Astrophysics and Heliophysics Science Divisions’ Research Program; and from the Virtual Planetary Laboratory Team, a member of the NASA Nexus for Exoplanet System Science, funded via NASA Astrobiology Program Grant No. 80NSSC18K0829.

\section*{Data Availability}

The outputs of our best YREC standard solar model and evolutionary history are available online. Other data are available upon request to the authors.


\bibliographystyle{mnras}
\bibliography{references} 



\appendix

\section{Climate Models}
\label{app:clima}
\begin{figure}
  \centering
	\includegraphics[width=\columnwidth]{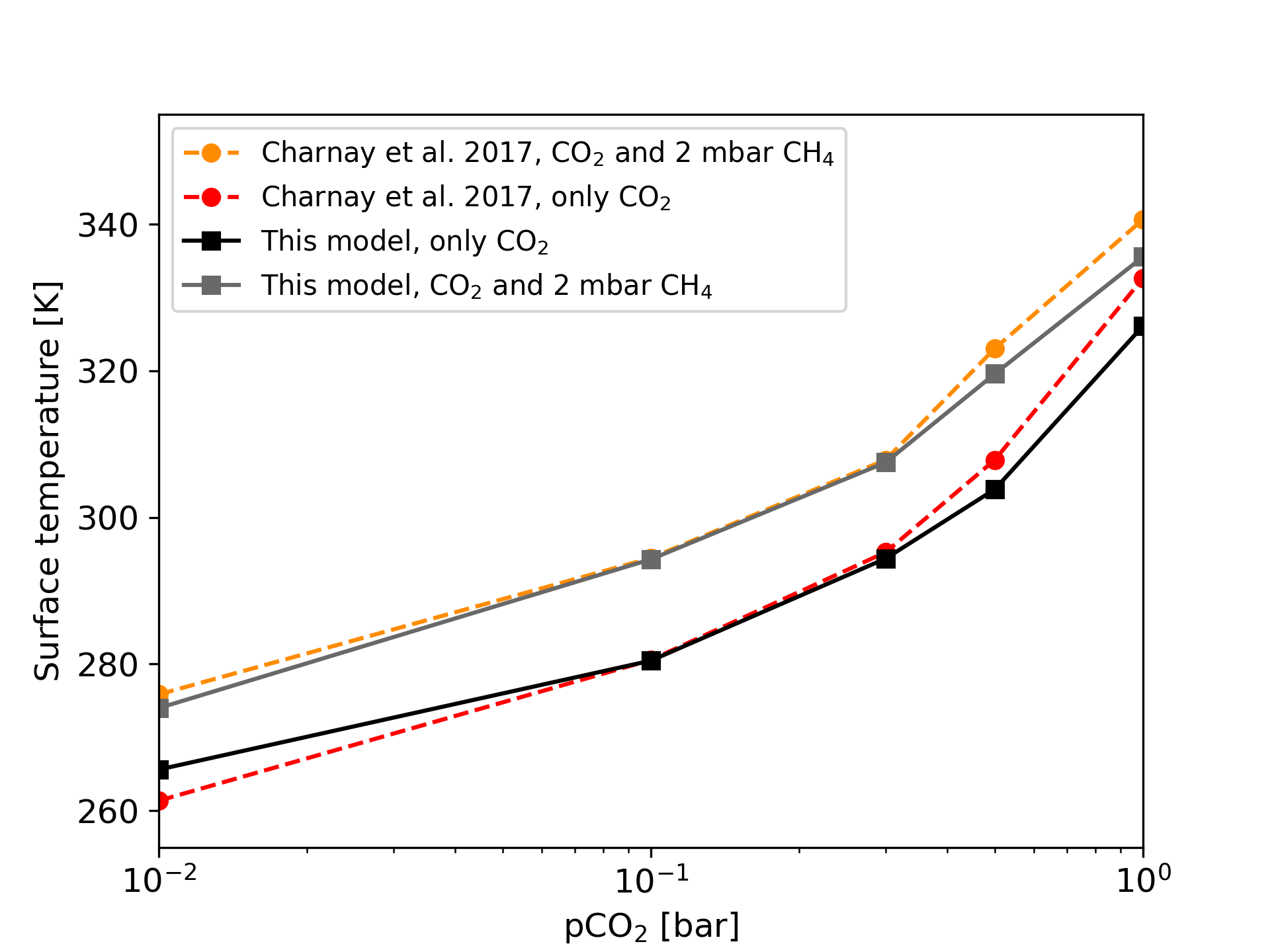}
    \caption{Comparison of forward 1-D climate model results with  \citet{Charnay2017} GCM global mean surface temperatures for 3.8 GA. All modeled atmospheres contain 1 bar N$_2$. }
    \label{fig:charnay_comp}
\end{figure}

We use \citet{Hayworth2020}'s version of the climate model originally developed by Kasting et al. In this version, the  $k$-coefficients for H$_2$O and CO$_2$ have been updated  with HITRAN2016 line list (\cite{GORDON2017}, see \citet{Vidaurri2022} for details).  The climate code does not include explicit clouds. Their effects on climate is accounted for implicitly by increasing the surface albedo to a higher value of 0.24 (compared to the modern value of $\sim$ 0.125), which we kept constant. This value was chosen as it reproduces modern Earth's  mean temperature of 288 K with modern atmospheric composition. Tropospheric water vapor was calculated according to the adapted version of the \citet{Manabe1967} parametrization- which is widely used in 1-D models- from \cite{Kasting1986}. Their adapted version is based on \citet{Cess1976} and allows the humidity in the upper troposphere to increase in response to higher temperatures. \\
For our inverse runs, a moist adiabat was used in the troposphere. In the stratosphere, the temperature profile was assumed to be isothermal at 171 K. This value was determined by running a set of 23 forward runs for solar constant values spanning 0.7-1.0 and CO$_2$ concentrations between 0.1-10 mbar as well as 20 mbar of. The inverse model was then run with the  the surface temperatures determined in the forward runs. The stratospheric temperatures were adjusted to reproduce the input solar constant to the fourth decimal with values ranging from 163-180 K, yielding an average of 171 K. \\
1-D climate models without explicit cloud representation tend to overestimate forcing by greenhouse gases and insolation changes (e.g. \cite{Fauchez2018}), and tend to overestimate greenhouse gas concentrations needed to maintain a temperate young Earth compared to more sophisticated GCMs (e.g. \cite{Charnay2020}). However, we note that our forward climate model reproduces the GCM results of \cite{Charnay2017} well enough for the questions addressed in this work (see Fig. \ref{fig:charnay_comp}).

\begin{table*}
\centering
\caption{Selected SSM properties in the trackfile. Horizontal lines separate rows of the header.}
\begin{tabular}{c c}
\multicolumn{1}{c}{Header}
&\multicolumn{1}{c}{Description}\\
\hline
Model \# & Model number in the run \\
AGE(Gyr) & Age of the star in Gyr \\
log(L/Lsun) & Luminosity of the star \\
log(R/Rsun) & Radius of the star \\
log(g) & Surface gravity of the star \\
log(Teff) & Effective temperature of the star \\
Mconv. env & Mass of the convective envelope \\
R & Radius at the base of the convection zone \\
T & Temperature at the base of the convection zone \\
Rho & Density at the base of the convection zone \\
P & Pressure at the base of the convection zone \\
kappa env & Opacity $\kappa$ at the base of the convection zone \\
\hline
Central: & \\
log(T) & Central temperature \\
log(RHO) & Central density \\
log(P) & Central pressure \\
X & Central hydrogen mass fraction \\
Y & Central helium mass fraction \\
Z & Central heavy element (metal) mass fraction \\
Luminosity: ppI & Energy generated by the p–p I branch \\
ppII & Energy generated by the p–p II branch\\
ppIII & Energy generated by the p–p III branch\\
CNO & Energy generated by the CNO cycle \\
\hline
Cl SNU & Calculated event rates in chlorine experiments \\
Ga SNU & Calculated event rates in gallium experiments \\
Neutrinos (1E10 erg/CM$^2$ at earth): & \\
pp, pep, hep, Be7, B8, N13, O15, F17 & Neutrino fluxes at Earth\\
\hline
Central Abundances: & \\
He3, C12, C13, N14, N15, O16, O17, O18 & Abundances of elements in the core \\
\hline
Surface Abundances: & \\
He3, C12, C13, N14, N15, O16, O17, O18, Li6, Li7, Be9 & Abundances of elements at the surface \\
X & Hydrogen mass fraction at the surface \\
Y & Helium mass fraction at the surface \\
Z & Heavy element (metal) mass fraction at the surface \\
Z/X & Heavy element (metal) to hydrogen ratio at the surface \\
\hline
Jtot & Total angular momentum \\
KE rot tot & Total rotational kinetic energy\\
total I & Total moment of inertia \\
CZ I & Moment of inertia of the convective envelope \\
Omega surf & Angular velocity of the convective envelope \\
Omega cent & Angular velocity of the interior of the star \\
Prot (day) & Rotation period of the star \\
Vrot (km/s) & Rotational velocity of the star \\
TauCZ (sec) & Convective overturn timescale \\

\end{tabular}
\label{tbl:track}
\end{table*}


\bsp	
\label{lastpage}
\end{document}